\documentclass{article}
\pdfoutput=1

\usepackage[text={6.75in,9.5in},centering,letterpaper]{geometry}
\usepackage{amssymb,amsmath,amsthm}
\usepackage{graphicx}
\usepackage[margin = 10pt, font=small, labelfont=bf, labelsep = endash]{caption}
\usepackage{subcaption}
\usepackage{xcolor}
\usepackage{bbm}

\title{\color{blue!30!black} Modelling honey bee colonies in winter using a Keller-Segel model with a sign-changing chemotactic coefficient\thanks{Last edited: \today.}}

\author{Robbin Bastiaansen\thanks{Mathematical Institute, Leiden University, 2300 RA Leiden, The Netherlands
  (r.bastiaansen@math.leidenuniv.nl, doelman@math.leidenuniv.nl, vivi@math.leidenuniv.nl)}
\and Arjen Doelman\footnotemark[2]
\and Frank van Langevelde\thanks{Resource Ecology Group, Wageningen University, P.O. Box 47, 6700 AA Wageningen, The Netherlands, and School of Life Sciences, Westville Campus, University of KwaZulu-Natal, Durban 4000, South Africa (frank.vanlangevelde@wur.nl)}
\and Vivi Rottsch\"afer\footnotemark[2]}

\newcommand{\R}{\ensuremath{\mathbb{R}}}
\DeclareMathOperator{\sech}{sech}
\newcommand{\arccosh}{\ensuremath{\mbox{arccosh}}}
\newcommand{\arctanh}{\ensuremath{\mbox{arctanh}}}

\newtheorem{remark}{\color{blue!30!black}Remark}

\begin{document}

\maketitle

\begin{abstract}
Thermoregulation in honey bee colonies during winter is thought to be self-organised. We added mortality of individual honey bees to an existing model of thermoregulation to account for elevated losses of bees that are reported worldwide. The aim of analysis is to obtain a better fundamental understanding of the consequences of individual mortality during winter. This model resembles the well-known Keller-Segel model. In contrast to the often studied Keller-Segel models, our model includes a chemotactic coefficient of which the sign can change as honey bees have a preferred temperature: when  the local temperature is too low, they move towards higher temperatures, whereas the opposite is true for too high temperatures. Our study shows that we can distinguish two states of the colony: one in which the colony size is above a certain critical number of bees in which the bees can keep the core temperature of the colony above the threshold temperature, and one in which the core temperature drops below the critical threshold and the mortality of the bees increases dramatically, leading to a sudden death of the colony. This model behaviour may explain the globally observed honey bee colony losses during winter.
\end{abstract}

\section{Introduction}

The reported global losses of honey bee colonies could have severe consequences for food production~\cite{gallai2009,klein2007,neumann2010}. Especially during winter, honey bee colonies experience high mortality of individual bees with the result that the colony goes extinct before the next spring season~\cite{hayes2008,dooremalen2012,doke2015}. For honey bees, the key to survive the winter period is the generation and preservation of heat. It has been demonstrated that colonies do not have a centralised mechanism to monitor and adjust in-hive temperature, and also that the thermoregulation does not depend on communication between bees~\cite{southwick1983,heinrich1981}. Instead, each individual honey bee has  a sensory-motor system that responds to external stimuli, such as local temperature differences in the colony, and thermoregulation during winter is therefore hypothesised to be self-organised~\cite{watmough1995self, heinrich1981, heinrich2013, moritz2012}. This self-organisation is the result of several processes. Firstly, bees produce heat through flight muscle activities~\cite{esch1960, esch1964}. Below a certain temperature the bee starts shivering with her flight muscles whereas above this temperature she remains at rest. Secondly, honey bees have a thermotactic movement which is based on temperature differences in their local neighbourhood~\cite{heinrich1981}. When a bee is too warm, she will move in the direction of lower temperature. If she is too cold, she will advance towards higher temperature. Based on these processes a model for the local bee density and the local temperature has been introduced in~\cite{watmough1995self}. However, in this model the mortality of individual bees is not included -- all bees stay alive surviving until the next spring season.  In this article, we formulate an extended model that does take the death of bees into account. Our model describes a honey bee colony during winter; no young bees emerge and are added to the colony. The aim of analysis in this article is to obtain a better fundamental understanding of the consequences of individual mortality of honey bees during winter.

The thermotactic movement of honey bees is dictated by local temperatures. The movement of organisms that arises in the direction of a gradient in the concentration of a substance -- often a chemical~\cite{keller1971,tindall2008} -- or in temperature has been observed frequently in nature. This process is generally called chemotaxis and has originally been described by Keller and Segel in the modelling of slime molds~\cite{keller1971}. A  generalisation of this model is given by the following two-component partial differential equation~\cite{hillen2009,tindall2008}:
\begin{equation}
	\begin{cases}
		\frac{\partial T}{\partial t} & = \Delta T + h(\rho,T), \\
		\frac{\partial \rho}{\partial t} & = \nabla \left[ \nabla \rho - \chi(T) \rho \nabla T \right] + g(\rho,T).
	\end{cases}\label{eq:generalisedKellerSegel}
\end{equation}
In this equation, $\rho$ is the density of the organisms and $T$ the concentration of the chemical (temperature in our case). The functions $h$ and $g$ specify the reaction terms -- creation/loss and growth/mortality. Evolution of the chemical $T$ arises from  diffusion and movement of the organisms $\rho$ by chemotaxis. The function $\chi$ is the so-called  chemotactic coefficient. When $\chi > 0$, movement is directed towards higher chemical concentrations; when $\chi < 0$, movement is away from it. Numerous studies have been conducted on this Keller-Segel model for various choices of functions $h$, $g$ and $\chi$ -- see for instance the review papers~\cite{tindall2008,hillen2009, bellomo2015} and references therein. The focus is often on global existence results or on finite-time blow-up~\cite{horstmann2003}. However, these studies focus on models with a positive chemotactic coefficient ($\chi > 0$) and no mortality of the organisms ($g \geq 0$). This is intrinsically different from the setting we consider here. In fact, we are not aware of studies of Keller-Segel models in the mathematical literature of the type considered in this article.

The thermotactic movement of honey bees is more subtle than can be described by taking a chemotactic coefficient $\chi>0$; bees do not always move towards the location with the highest temperature, but they have  a preferred temperature $T_\chi$. This is reflected in their movement: when  the local temperature $T$ is too low, $T < T_\chi$, they move towards higher temperatures; when $T > T_\chi$, they move away to lower temperatures. This means that the chemotactic coefficient $\chi(T)$ changes sign at $T_\chi$, and hence it can become negative, which is very different from the generalised Keller-Segel models~\eqref{eq:generalisedKellerSegel} where $\chi$ has a fixed, positive sign.
Moreover, to be able to study bee losses, we need to incorporate mortality of the bees. This leads to the following model
\begin{equation}
	\begin{cases}
		\frac{\partial T}{\partial t} & = \Delta T + f(T)\rho, \\
		\frac{\partial \rho}{\partial t} & = \nabla \left[ \nabla \rho - \chi(T) \rho \nabla T \right] - \theta(\rho,T)\rho,		
	\end{cases}\label{eq:KellerSegelForBees}
\end{equation}
where $\rho\geq 0$ is the bee density and $T$ the local temperature. Although our model still has the structure of~\eqref{eq:generalisedKellerSegel} (where $h(\rho,T) = f(T)\rho$ models the heat generation by bees and $g(\rho,T) = \theta(\rho,T)\rho$ the individual mortality of bees), our setting, with $\chi$ changing sign from positive to negative, generates very different dynamics compared to the models considered in the mathematical literature. This model is an extension of the bee model in~\cite{watmough1995self}, which also includes a chemotactic coefficient $\chi$ that changes sign. To that model, we have added bee mortality $\theta(\rho,T) > 0$. On the other hand, \eqref{eq:KellerSegelForBees} also is a simplification of the model introduced in~\cite{watmough1995self}. For instance, we assume that the diffusion coefficients are constant which is different from~\cite{watmough1995self} where these are functions of $\rho$. This is not a major modification since the functions describing diffusion in~\cite{watmough1995self} -- which come from observations -- are indeed almost constant. 

In~\cite{watmough1995self}, the functions $f$ and $\chi$ have also been based on observations. In our analysis, we found that the precise form of the functions $f$ and $\chi$ does not alter the qualitative aspects of the results. To clarify the  presentation and to enable explicit asymptotic analysis, we have simplified these functions, based on those in~\cite{watmough1995self}. As previously discussed, it is important that the function $\chi$ switches signs from $\chi(T) > 0$ for small $T$ to $\chi(T) < 0$ for large $T$. Therefore, we choose $\chi$ to be a step-function; similarly, based on the data in~\cite{watmough1995self} $f$ is chosen to be a step-function as well.
\begin{equation}
	f(T) = \begin{cases} f_{\mathrm{low}}, & T < T_f \\ f_\mathrm{high}, & T > T_f\end{cases}; \hspace{1cm} \chi(T) = \begin{cases} +\ \chi_1, & T < T_\chi; \\ -\ \chi_2, & T > T_\chi, \end{cases}\label{eq:definitionchi}
\end{equation}
where $f_\mathrm{low},f_\mathrm{high}, \chi_1, \chi_2 > 0$, $T_f$ is the temperature where $f$ changes value and $T_{\chi}$ is the temperature where $\chi(T)$ changes sign ($T_f < T_\chi$). The temperature $T_\chi$ can be thought of as the preferred temperature for the bees, as bees prefer to move toward locations with this temperature. 

In nature, honey bees form combs for brood and storage of honey and pollen (which are also offered by beekeepers in hives) and bees are found to cluster in between the combs (i.e. the inter-comb spaces), with the highest temperature in the centre~\cite{stabentheiner2003,stabentheiner2010}. For the modelling, we take a cross-section of this cluster from the centre to the edge, and thus study the cluster in one spatial dimension (but without going explicitly into polar coordinates -- for simplicity). Therefore, we analyse the model on $[0,L]$ where $x = 0$ is the centre of the colony and $x = L$ is the edge of it. 

The above equations  need to be completed by a set of four boundary conditions. Bees do not leave the colony in winter, and therefore, we  impose no-flux boundary conditions at $x = L$, leading to a (bee) mass conservation in the system when mortality is absent. Also, the temperature at the edge of the colony  at $x=L$ is equal to the  ambient temperature $T_a$, which we assume to be fixed and to be below the preferred temperature $T_\chi$ in winter. Finally, because of the assumed symmetry we need to impose boundary conditions for both $T$ and $\rho$ at the centre of colony. 

Summarizing, the complete model is given by
\begin{equation}
	\begin{cases}
		\frac{\partial T}{dt} & = \frac{\partial^2 T}{\partial x^2} + f(T) \rho, \\
		\frac{\partial \rho}{dt} & = \frac{\partial^2 \rho}{\partial x^2} - \frac{\partial}{\partial x} \left[ \chi(T) \rho \frac{\partial T}{\partial x} \right] -  \theta(\rho,T) \rho.
	\end{cases}\label{eq:beePDE}
\end{equation}
\begin{align}
	T_x(0,t) & = 0; & \rho_x(0,t) & = 0; \label{eq:boundc1}\\
	T(L,t) & = T_a<T_{\chi}; & \left( \rho_x - \chi(T) \rho T_x \right) (L,t) & = 0,
\label{eq:boundc2}
\end{align}
where the subscripts $x$ denote derivatives with respect to $x$.

\begin{remark}
In~\cite{watmough1995self}, the boundary of the honey bee colony is not fixed at some value $L$ but is a free boundary, i.e. it is allowed to move in time. This leads to more complex conditions at the boundary, see section \ref{sec:movingBoundary}.
\end{remark}

The novel modelling aspect of the present work is the inclusion of a nontrivial mortality rate $\theta$; only the model for $\theta=0$ has been studied before. We formulate a mortality coefficient $\theta$ based on observations. During winter, bees die from aging, and therefore, the mortality of bees is highly influenced by the amount of work a bee has to perform; moreover, it is amplified by parasites such as the mite {\it Varroa destructor}~\cite{amdam2002,amdam2009,dooremalen2012}. We postulate that there are three distinctive effects that contribute to individual mortality: (i) the effect of the local temperature ($\theta_T$); (ii) the effect of the length of the resting times between bouts of heat generation, which is closely related to the effective refresh rates of heat generating bees ($\theta_D$); (iii) the effect of parasitic mites in the colony ($\theta_M$). The effective mortality coefficient is then given by the product of these effects, i.e. 
\begin{equation}
\theta(T,\rho) = \theta_0 \theta_T(T)\ \theta_D(\rho)\ \theta_M(\rho), 
\label{eq:mort}
\end{equation}
where $\theta_0$ is a constant  that needs to be tuned to align with observations.

The first effect, of temperature, represents that mortality does not increase when the (local) temperature is above a certain threshold, $T_\theta > T_a$. If the temperature is too low, i.e. $T(x) < T_\theta$, a bee in that location has to work (too) hard to generate heat, reducing her lifespan. Mathematically we, once again, strongly simplify this effect and describe this by the step-function
\begin{equation}
\theta_T(T) =
	\begin{cases}
		1, & \mbox{if $T < T_\theta$};\\
		0, & \mbox{if $T \geq T_\theta$}.
	\end{cases}
\end{equation}

The second effect comes from the ratio between local bee density $\rho$ and colony size $\rho_\mathrm{tot}$ which is called the refresh rate by recovered bees. To heat up the colony bees work together; each bee can generate heat by shivering her flight muscles, but only for around 30 minutes after which she needs to recover and refill reserves by consuming honey~\cite{tautz2008}. Therefore, at each moment, bees at the periphery of the colony can become (too) cold and have to work hard to generate heat, while bees inside the colony, at warmer locations, rest and recover from earlier heat generation bouts before starting to generate heat again. After a while, the bees rotate so that recovered bees can take over the heating duty, enabling the heating bees to recover~\cite{stabentheiner2003}. Therefore, if there are a lot of bees in the colony, bees can rest and recover longer between heating bouts. If the colony is relatively small and the opportunities to recover are short, bees may have to work more frequently reducing their lifespan. Mathematically, the contribution of the refresh rate by recovered bees to the mortality is modelled as

\begin{equation}
	\theta_D(\rho) = \frac{\rho}{(\rho_\mathrm{tot})^\gamma}, \hspace{1cm} (\gamma > 0)
\end{equation}
where $\gamma > 0$ is some unknown exponent.

The third contribution stems from the (currently excessive) presence of the parasitic mite {\it Varroa destructor} in honey bee colonies~\cite{dooremalen2012}. Although there is a general agreement that there is no single
explanation for the extensive colony losses, and that interactions
between different stresses are likely to be involved, the presence of
{\it V. destructor} in colonies places an important pressure on bee health. {\it V. destructor} reduces the body weight and protein content of individual bees, which is found to shorten their lifespan~\cite{vandooremalen2013, amdam2004}. Thus, if the amount of mites per bee increases, bee mortality increases as well. Because mites may jump to neighbouring bees when their host bee dies, this fraction increases when colony size decreases. Mathematically, we model this effect as
\begin{equation}
	\theta_M(\rho) = 1 + \frac{m}{\rho_\mathrm{tot}},
\end{equation}
where $m$ is the amount of mites present in the colony.

\begin{figure}
	\centering
	\begin{subfigure}[t]{0.475 \textwidth}
		\centering
		\includegraphics[width = \textwidth]{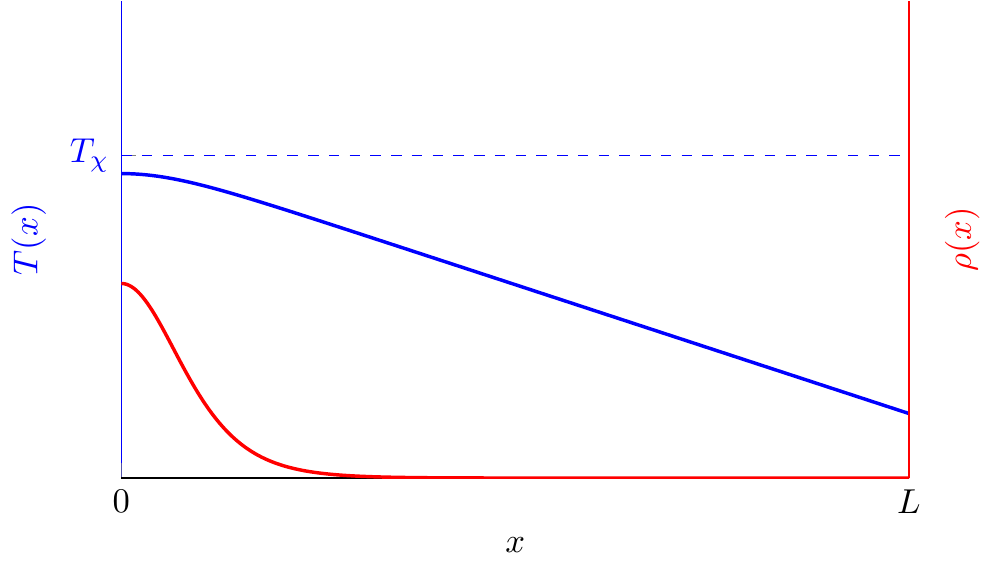}
		\caption{$\rho(x)$ and $T(x)$ for a type I solution}
		\label{fig:typeIsolFORM}
	\end{subfigure}
~
	\begin{subfigure}[t]{0.475 \textwidth}
		\centering
		\includegraphics[width = \textwidth]{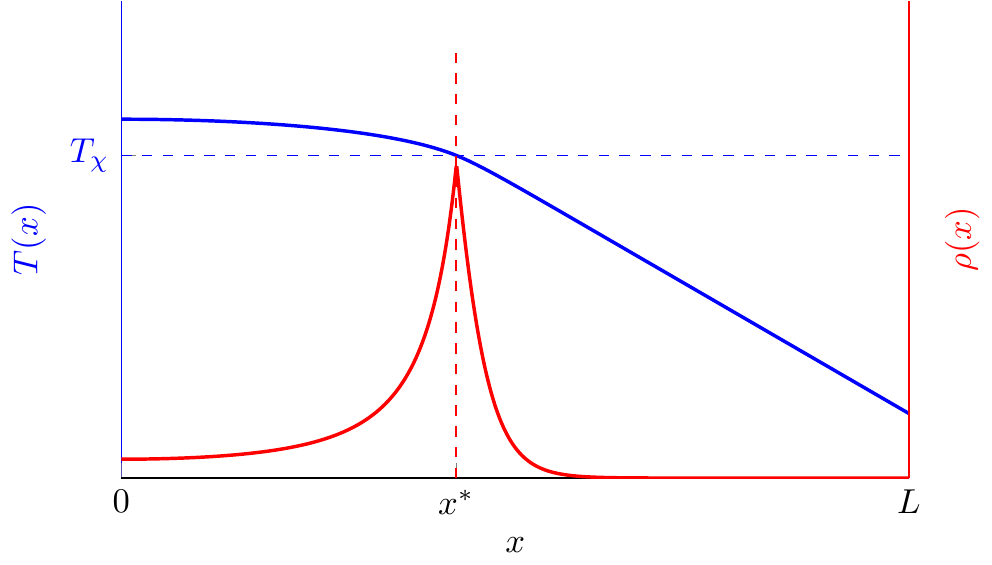}
		\caption{$\rho(x)$ and $T(x)$ for a type II solution}
		\label{fig:typeIIsolFORM}
	\end{subfigure}
\caption{Plots of a type I solution (a) and a type II solution (b). The bee density $\rho(x)$ has a maximum at $x = x^*$, the location where $T(x^*) = T_\chi$, in the type II solutions. When $\rho_\mathrm{tot} < \rho_{\mathrm{tot},c}$ a type I solution is obtained and when $\rho_\mathrm{tot} > \rho_{\mathrm{tot},c}$ a type II solution is found.}
\label{fig:intro-steadyStateTypes}
\end{figure}

In this article, we study the model in~\eqref{eq:beePDE} with boundary conditions~\eqref{eq:boundc1}-\eqref{eq:boundc2}. First, we analyse system~\eqref{eq:beePDE} in the absence of bee mortality, so we set $\theta=0$. Under that assumption, we study the steady state solutions of the model in section~\ref{sec:steadyStates}. There, we find two types of steady state configurations, type I and type II, distinguishable by the colony size, $\rho_\mathrm{tot} = \int_0^L \rho(x)\ dx$. We find that there exists a critical colony size $\rho_{\mathrm{tot},c}$ such that steady states are of type I -- see Figure~\ref{fig:typeIsolFORM} -- when $\rho_\mathrm{tot}<\rho_{\mathrm{tot},c}$, and of type II -- see Figure~\ref{fig:typeIIsolFORM} -- when $\rho_\mathrm{tot}>\rho_{\mathrm{tot},c}$. When the colony size is below the critical value $\rho_{\mathrm{tot},c}$, there are insufficient bees in the colony to increase the core temperature above  the bees' preferred temperature $T_\chi$. Therefore, bees cluster together at the colony's core and the corresponding  steady states  are denoted by type I -- see Figure~\ref{fig:typeIsolFORM} for a side view of this steady state. On the other hand, when the colony size lies above the critical size $\rho_{\mathrm{tot},c}$, there are enough bees to keep the colony's core temperature above the bees' preferred temperature $T_\chi$. As a result, the distribution of bees has a peak at the location with temperature $T_\chi$. This steady state is denoted as a type II solution in this article; a side view of this configuration is given in Figure~\ref{fig:typeIIsolFORM}. The latter type of bee distribution, with bees clustering near the edge of the colony, is also observed in real honey bee colonies~\cite{southwick1971}.

Moreover, the simplicity of the model enables us to derive analytic expressions for these two types of steady state solutions in section~\ref{sec:steadyStates}. There, we find that it is  possible to obtain a closed form expression for the critical colony size $\rho_{\mathrm{tot},c}$ that forms the threshold between the two types of steady state solutions -- see equation~\eqref{eq:criticalbiomass}. We find that $\rho_{\mathrm{tot},c}$ increases (almost linearly) when the ambient temperature $T_a$ decreases. Thus, a larger colony is needed to adequately heat the colony when it is colder.

In section~\ref{sec:movingBoundary} we explore the effects of the addition of a moving boundary to the model -- like in the original model in~\cite{watmough1995self}. This leads to a more complex model formulation. However, the steady state analysis is not altered qualitatively; again the same two types of solutions exist which are again distinguishable by the total amount of bees in the colony.

Subsequently, in sections~\ref{sec:mortality} and~\ref{sec:simulations}, we take the mortality of individual bees into account, hence the colony size decreases. Compared to their movement, the mortality of bees takes place on a much slower time scale. Therefore the solutions closely follow the steady state configurations, with the (decreasing) $\rho_\mathrm{tot}$ acting as the slowly varying parameter. We choose realistic values for the parameters and  study the evolution of $\rho_\mathrm{tot}$ using numerical simulations. These simulations reveal a speed-up in  decrease of the colony size when $\rho_\mathrm{tot}$ decreased below $\rho_{\mathrm{tot},c}$, see Figure~\ref{fig:simulationDataTa} -- for a type I solution the colony goes extinct very quickly. Therefore, the survival a colony is increased when it remains sufficiently long in a type II configuration. Here, sufficiently long in practice means beyond the end of the winter season: a colony survives the winter if it succeeds in remaining of type II until the beginning of the next spring season (when the bees leave the colony to forage and young bees are produced). As simulations show, bee colonies remain of type II for a longer period when there are (i) less mites, (ii) higher ambient temperatures and (iii) a larger initial colony size (at the start of winter). In section~\ref{sec:simulations}, we give details  of the precise effects on the decline of $\rho_\mathrm{tot}$ over a winter period of the three mortality parameters mentioned, and we also discuss their impact on the survival of the colony.

As a short encore, in section~\ref{sec:simulations-multipleRows}, we consider the possibility of multiple combs in a hive so that the colony is divided in parts occupying several inter-comb spaces, which are connecting by moving bees going from the one inter-comb space to the other (through the comb or going around the comb). For this, we present a simple extension of~\eqref{eq:beePDE} that takes into account multiple inter-comb spaces and the movement of bees between them. Using simulations we show how having multiple combs is beneficial for the survival of a bee colony.

Finally, we briefly discuss the implications of our findings and indicate future lines of research in the concluding section~\ref{sec:conclusions}.

\begin{remark}\label{energyFunctionalRemark}
When $\rho_\mathrm{tot} < \rho_{\mathrm{tot},c}$ the steady state configuration is of type I. For these solutions, $T < T_\chi$ over the whole domain and therefore $\chi > 0$ everywhere. Hence, these type of solutions are essentially covered by the classical Keller-Segel theory. Of specific interest is the existence of an energy functional~\cite{blanchet2015}
\begin{equation}
	\mathcal{E}[\rho,T] := \int_0^L \left( \frac{1}{\chi} \rho \log \rho - T \rho + \frac{1}{2} T_x^2 \right)\ dx,
	\label{eq:energyFunctional}
\end{equation}
which is bounded from below and ever-decreasing when $\chi > 0$ everywhere~\cite{blanchet2015}. From this it can be deduced that type I configurations are stable solutions to~\eqref{eq:beePDE} in the absence of mortality. We are not aware of a similar argument for type II solutions, when $\chi$ does change sign (since $T(0) > T_\chi$ and $T(L) = T_a < T_\chi$). At least, it can be checked that the energy functional~\eqref{eq:energyFunctional} does no longer suffice, as the energy can grow under these conditions.
\end{remark}

\section{Steady States}\label{sec:steadyStates}
In this section, we first study stationary solutions of system~\eqref{eq:beePDE} without mortality, i.e. $\theta(\rho,T) \equiv 0$. For clarity of presentation we also additional take $f_\mathrm{low} = f_\mathrm{high} = \bar{f}$ first; at the end of the section we comment on the additional effects in case $f_\mathrm{low} \neq f_{\mathrm{high}}$. Then, steady state configurations $(T(x),\rho(x))$ of~\eqref{eq:beePDE} must satisfy
\begin{equation}
	\begin{cases}
		 T_{xx} + f(T) \rho_s & =0, \\
		 \rho_{xx} - (\chi(T)\rho_s T_{x})_x & =0.
	\end{cases}
\end{equation}
The second equation can be integrated from $0$ to $x$ and because of the no-flux boundary conditions for the local bee density~\eqref{eq:boundc2}, we find
\begin{equation}
	\begin{cases}
		 T_{xx} + f(T) \rho & =0, \\
		 \rho_{x} - \chi(T)\rho T_{x}& =0.
\label{eqint}
	\end{cases}
\end{equation}
Since $f(T) > 0$ for all $T \in \R$, it follows from the first equation that
\begin{equation}
	\rho = - T_{xx} / f(T).
	\label{eq:rhosrelation}
\end{equation}
Substituting this into the second equation of~\eqref{eqint} and multiplying the result with $f(T)$ leads to
\begin{equation}
- T_{xxx} + \left( \frac{f'(T)}{f(T)} + \chi(T)\right) T_{x} T_{xx}=0.
\end{equation}
 Introducing $S := T_{x}$ and $R := S_x$ yields
\begin{equation}
	\begin{cases}
		T_{x} & = S, \\
		S_x & = R,\\
		R_x & = \left( \chi(T) + \frac{f'(T)}{f(T)} \right) S R.
	\end{cases}
	\label{eq:steadystateODEgen}
\end{equation}
Now, since $f(T) \equiv \bar{f}$, this system reduces to
\begin{equation}
	\begin{cases}
		T_{x} & = S, \\
		S_x & = R,\\
		R_x & =  \chi(T) S R.
	\end{cases}
	\label{eq:steadystateODE}
\end{equation}
Note that since the local bee density satisfies  $\rho\geq 0$, it follows from condition~\eqref{eq:rhosrelation}
that $R=T_{xx}\leq 0$.

\begin{figure}[t!]
	\centering
	\begin{subfigure}[t]{0.4 \textwidth}
		\includegraphics[width = \textwidth]{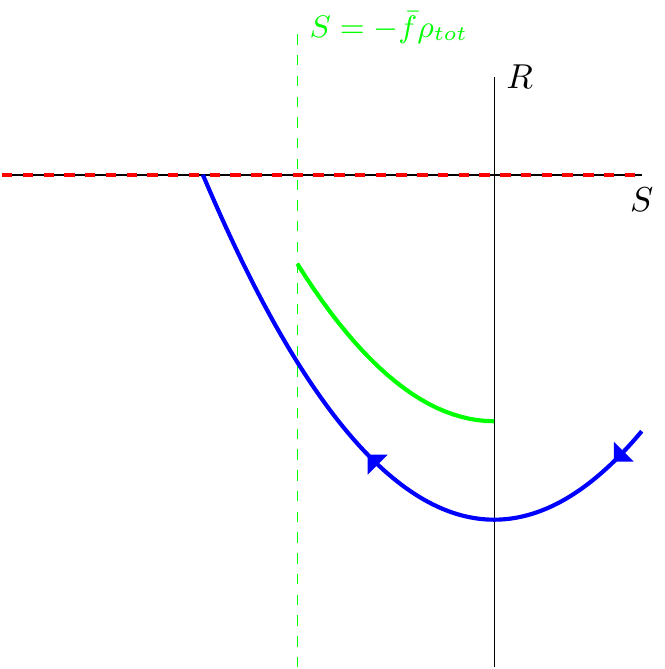}
		\caption{$\bar{\chi} > 0$}
		\label{fig:RSpplanePOS}
	\end{subfigure}
~
	\begin{subfigure}[t]{0.4 \textwidth}
		\includegraphics[width = \textwidth]{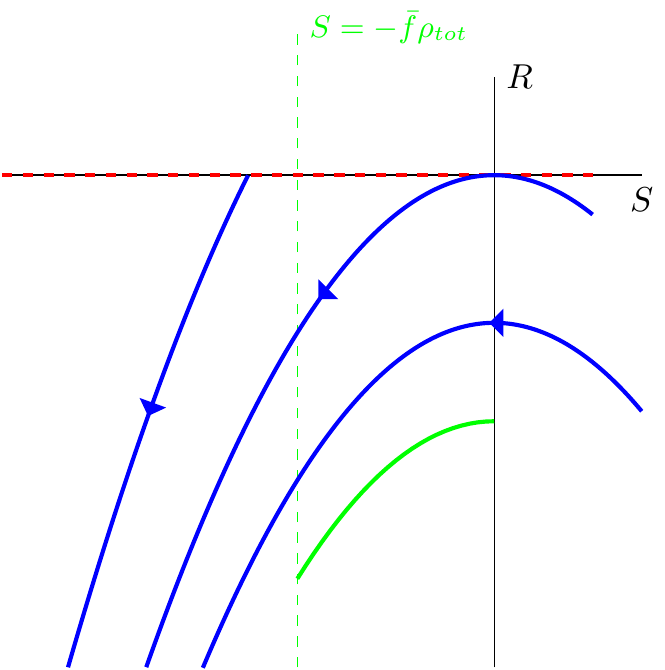}
		\caption{$\bar{\chi} < 0$}
		\label{fig:RSpplaneNEG}
	\end{subfigure}
	\caption{Sketches of the $(R,S)$ phase plane for the ODE in~\eqref{eq:steadystateODE} in case of a constant $\chi$, i.e. with $\chi(T) \equiv \bar{\chi}$. The dashed red lines indicate the line $\{R = 0\}$ on which both $S_x = 0$ and $R_x = 0$. The dashed green line is the line $\{S = -\bar{f} \rho_\mathrm{tot}\}$ and the solid green line indicates a solution that satisfies the boundary conditions $S(0) = 0$ and $S(L) = - \bar{f} \rho_\mathrm{tot}$. }
	\label{fig:RSpplane}
\end{figure}

This system needs to be accompanied by three boundary conditions. Naturally, the steady state solution should satisfy the boundary conditions of the original PDE. However, these only give us two boundary conditions, since all solutions of  system~\eqref{eq:steadystateODE} satisfy the no-flux boundary condition at $x=L$ for $\rho$  automatically.
This no-flux boundary condition does imply that the colony size in the domain, $\rho_\mathrm{tot} := \int_0^L \rho(x) dx$, needs to remain the same, i.e. mass conservation. This leads to a third boundary condition by using the expression for $\rho$ in terms of  $T_{xx}$ in equation~\eqref{eq:rhosrelation} -- given a constant $f$. Substituting~\eqref{eq:rhosrelation} into $\rho_\mathrm{tot}$ leads to the relation $S(L) - S(0) = - \rho_\mathrm{tot} \bar{f}$. To summarise, the boundary conditions for solutions of~\eqref{eq:steadystateODE} are
\begin{equation}\label{eq:boundaryConditionsODE}
	T(L) = T_a \hspace{0.2cm} \mbox{ (where } T_a<T_{\chi}\mbox{)}; \hspace{2cm}     S(0) = 0; \hspace{2cm} S(L) = - \bar{f} \rho_\mathrm{tot}.
\end{equation}
Note that this last condition introduces the non-trivial impact of $\rho_\mathrm{tot}$ on the nature of the steady state solutions.

From the fact that the function $\chi(T)$ is a step function, it follows that  the dynamics of solutions of~\eqref{eq:steadystateODE} are governed by the 2D subsystem of $S$ and $R$. If we for a moment assume that $\chi(T) \equiv \bar{\chi}$ is constant, we find, depending on the sign of $\bar{\chi}$, two qualitatively different ODEs, see Figure~\ref{fig:RSpplane}(a) for a  sketch of the $(R, S)$-plane for $\bar{\chi}>0$ and Figure~\ref{fig:RSpplane}(b) for $\bar{\chi}<0$. Here only the lower part of the plane is relevant since $R \leq 0$.

From the boundary condition $S(0) = 0$ we find  that at $x = 0$ the solution starts on the half-line $\{S = 0, R \leq 0\}$. On this half-line the flow of the ODE dictates that $S_x = R \leq 0$, and therefore, the solution is contained in the region $\{ S \leq 0, R \leq 0 \}$. The other boundary condition, $S(L) = -\bar{f} \rho_\mathrm{tot}$, indicates that the solution must end at $x = L$, on the half-line $\{S = - \bar{f} \rho_\mathrm{tot}, R \leq 0\}$, the dashed green lines in Figure~\ref{fig:RSpplane}. In general, for a fixed $L$ only one solution satisfies these constraints; in Figure~\ref{fig:RSpplane} we have sketched these solutions for both $\bar{\chi} > 0$ and $\bar{\chi} < 0$.

However, in our model $\chi$ is a piecewise constant function that changes sign at $T = T_\chi$, see~\eqref{eq:definitionchi}. Thus if $T < T_\chi$, the chemotactic constant $\chi(T)$ is positive and the phase portrait is as in Figure~\ref{fig:RSpplanePOS}; for $T > T_\chi$, $\chi(T)$ is negative and its phase portrait is given in Figure~\ref{fig:RSpplaneNEG}. Next, we construct solutions by combining  both phase planes in Figure~\ref{fig:RSpplane}. 

\begin{figure}[t!]
	\centering
	\begin{subfigure}[t]{0.475 \textwidth}
		\centering
		\includegraphics[width = 0.9\textwidth]{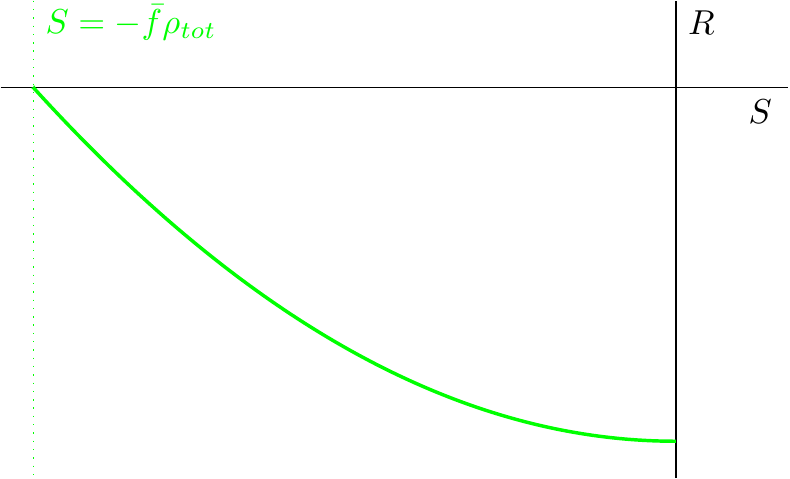}
		\caption{$(R,S)$-plane for a type I solution}
		\label{fig:RSPlaneTypeI}
	\end{subfigure}
~
	\begin{subfigure}[t]{0.475 \textwidth}
		\centering
		\includegraphics[width = 0.9\textwidth]{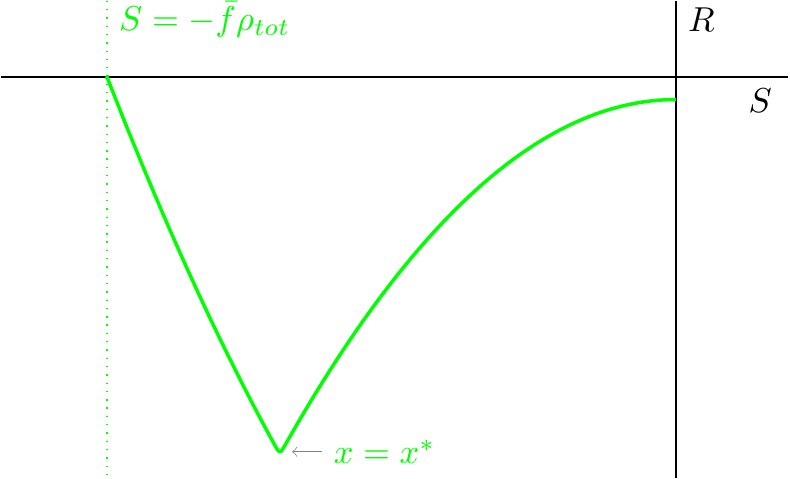}
		\caption{$(R,S)$-plane for a type II solution}
		\label{fig:RSPlaneTypeII}
	\end{subfigure}
\caption{Plot of the phase plane for a type I solution (a) and a type II solution (b). The corresponding plots of $T(x)$ and $\rho(x)$ can be found in Figure~\ref{fig:intro-steadyStateTypes}.}
\label{fig:steadystatesolutions}
\end{figure}

The boundary condition $T(L) = T_a < T_\chi$ ensures that close to the colony's edge $\chi(T) > 0$. Moreover, from the fact that  $S(x) \leq 0$ for all $x \in [0,L]$, and $T_x=S$,  we know that the temperature is decreasing (or constant if $\rho_\mathrm{tot} = 0$). This results in  two possible scenarios that we denote by type I and type II,  depending on the heat production of the bees:
\begin{itemize}
	\item[I:] The temperature stays below $T_\chi$ in the whole colony; i.e. $T(x) < T_\chi$ for all $x \in [0,L]$;
	\item[II:] The temperature is larger than
 $T_\chi$ at $x=0$, and hence, 
 there exists a point $x^*$ such that $T(x) > T_{\chi}$ for $x < x^*$ and $T(x) < T_{\chi}$ for $x > x^*$.
\end{itemize}
Both cases lead to different forms of steady state solutions. In the first situation the solution is described by system \eqref{eq:steadystateODE} with $\bar{\chi} > 0$ for all $x$, see Figure~\ref{fig:RSPlaneTypeI} for a sketch of the solution in the $(R,S)$-plane. However, in the second situation the solution first follows the solution in the phase plane for $\bar{\chi} < 0$ and then switches  at $x = x^*$ to the phase plane for $\bar{\chi} > 0$, see Figure~\ref{fig:RSPlaneTypeII} for the combined phase plane. The corresponding solutions $T$ (in blue) and $\rho$ (in red) for the type I solutions are plotted in Figure~\ref{fig:typeIsolFORM} and for the type II solutions in Figure~\ref{fig:typeIIsolFORM}.

The clear distinction between the two types of solutions is the absence/presence of a peak in the local bee density $\rho$. Observe that bees cluster at the centre of the colony in type I solutions, whereas for a type II solution they form a band at some location $x = x^*$. It has been observed  that bees in colonies also form these bands~\cite{southwick1971}.

\subsection{Explicit expressions for the steady state solutions}

In the previous section we qualitatively found two types of steady state configurations. We can also determine closed form solutions since system~\eqref{eq:steadystateODE} can be solved explicitly. In this section we determine these expressions and use them to find a criterion to distinguish between the two types; we show that there exists a critical colony size $\rho_{\mathrm{tot},c}$ at which the solution type changes: when  $\rho_\mathrm{tot} < \rho_{\mathrm{tot},c}$ then the steady state solution is of type I and when $\rho_\mathrm{tot} > \rho_{\mathrm{tot},c}$ it is of type II.

\subsubsection{Type I solutions}

\begin{figure}
	\centering
	\begin{subfigure}[t]{0.475 \textwidth}
		\centering
		\includegraphics[width = \textwidth]{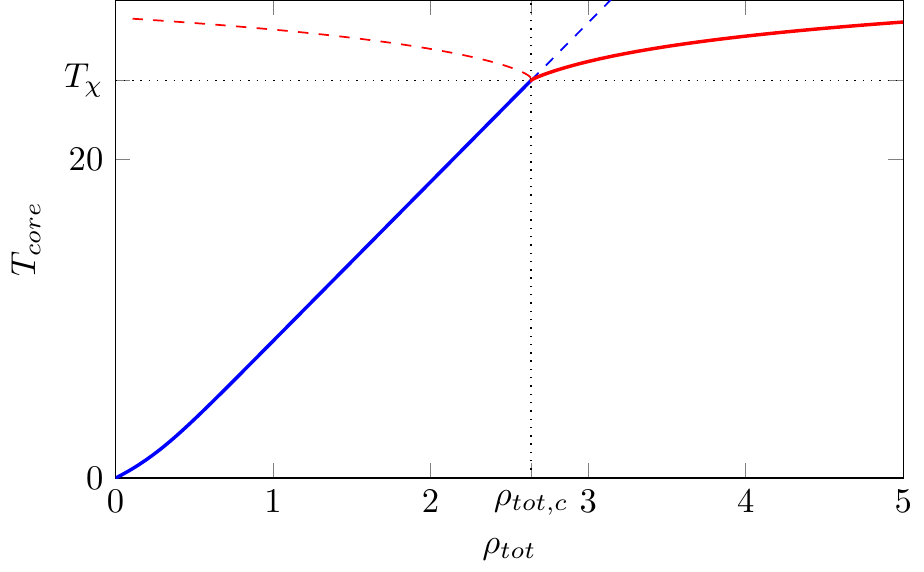}
		\caption{$T_\mathrm{core}$ as function of $\rho_\mathrm{tot}$.}
	\end{subfigure}
~
	\begin{subfigure}[t]{0.475\textwidth}
		\centering
		\includegraphics[width=\textwidth]{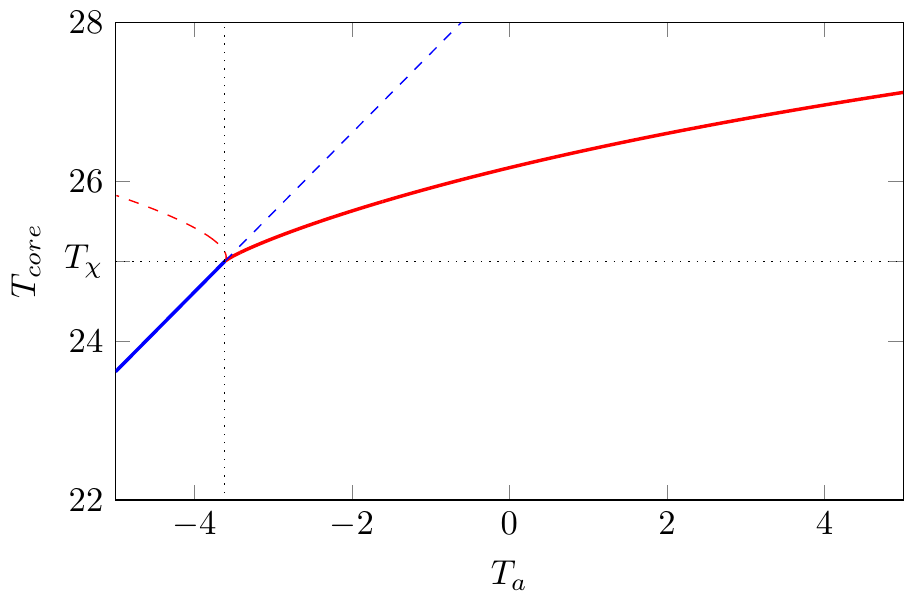}
		\caption{$T_\mathrm{core}$ as function of $T_a$.}
	\end{subfigure}
	\caption{Plot of the core temperature $T_\mathrm{core}$ as function of the colony size $\rho_\mathrm{tot}$ (a) and the ambient temperature $T_a$ (b). The blue lines indicate the core temperature for type I solutions, whereas the red lines indicate the core temperature for a type II solution. The dashed lines indicate solutions that are not realistic. Parameter values used are $\bar{f} = 1$, $\chi_1 = 1$, $\chi_2 = 1$, $L = 10$, $T_\chi = 25$ and $T_a = 0$ (a) or $\rho_\mathrm{tot} = 3$ (b).}
	\label{fig:TcorePlots}
\end{figure}

 In type I solutions, the temperature does not exceed $T_\chi$, and  therefore, $\chi(T) = \chi_1 > 0$ for all $x \in [0,L]$. Then, the last two equations in system~\eqref{eq:steadystateODE}, for $S$ and $R$,  can be written as a second order differential equation in $S$,
\begin{equation}
	0 = S_{xx} - \chi_1 S S_x.
	\label{eq:typeIODE}
\end{equation}
This equation can be integrated  once and  yields the first order ODE,
\begin{equation}
	S_x = \frac{\chi_1}{2} \left(S^2 -  C_1\right),
\end{equation}
where $C_1$ is an integration constant that is not yet determined.
This constant $C_1$ must be positive since we know that 
$S_x \leq 0$ for all $x$. 
 The solution to the above  equation is given by
\begin{equation}
	S(x) = - \sqrt{C_1} \tanh\left( \frac{\chi_1}{2} \sqrt{C_1} x + C_2\right),
\label{eq:SsolI}
\end{equation}
where $C_2$ is another constant. The boundary conditions for $S$ in~\eqref{eq:boundaryConditionsODE} imply that $C_2 = 0$, while $C_1$ must satisfy
\begin{equation}
	\sqrt{C_1} \tanh\left( \frac{\chi_1}{2} \sqrt{C_1} L \right) = \bar{f} \rho_\mathrm{tot}.
	\label{eq:typeIC1general}
\end{equation}
We obtain the temperature profile by integrating the expression for $S$ once;
\begin{equation}
	T(x) = T_\mathrm{core} - \frac{2}{\chi_1} \log\left[ \cosh\left( \frac{\chi_1}{2} \sqrt{C_1} x \right)\right],
\end{equation}
where $T_\mathrm{core}$ is the temperature  at $x = 0$. With the last boundary condition, $T(L) = T_a$, the temperature $T_\mathrm{core}$  is determined as
\begin{equation}
	T_\mathrm{core} = T_a + \frac{2}{\chi_1} \log\left[ \cosh\left( \frac{\chi_1}{2} \sqrt{C_1} L \right)\right].
	\label{eq:typeITcore}
\end{equation}
In Figure~\ref{fig:TcorePlots}(a), we plot $T_\mathrm{core}$ as a function of $\rho_\mathrm{tot}$ by combining~\eqref{eq:typeITcore} and~\eqref{eq:typeIC1general}.

Finally, $\rho$ can be determined from~\eqref{eq:rhosrelation} yielding 
\begin{equation}
\rho = - T_{xx} / \bar{f}=- S_{x} / \bar{f} = \frac{C_1\chi_1}{2\bar{f} \left[ \cosh\left( \frac{\chi_1}{2} \sqrt{C_1} x \right)\right]^2}. \label{eq:rhoDistributionTypeI}
\end{equation}
A plot of this steady state solution is given in Figure~\ref{fig:typeIsolFORM}. 

\begin{figure}
	\centering
	\begin{subfigure}[t]{0.475\textwidth}
	\centering
	\includegraphics[width=\textwidth]{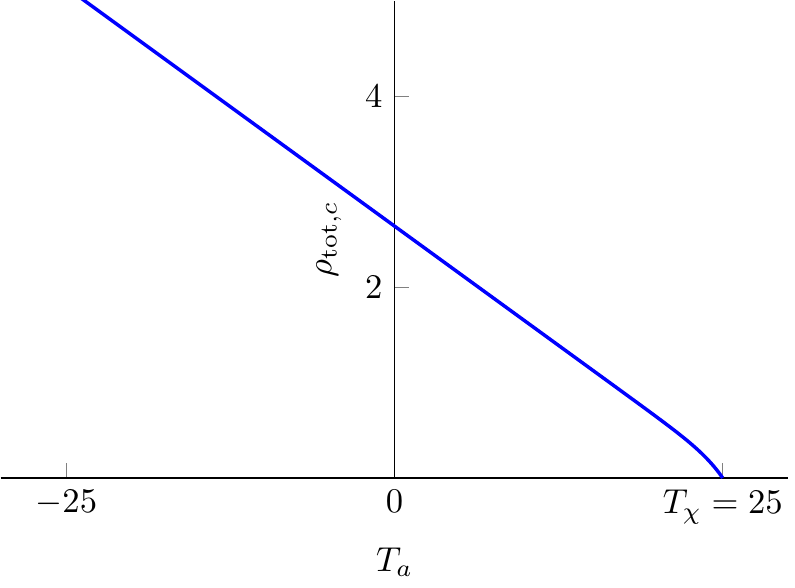}
	\caption{$\rho_{\mathrm{tot},c}$ as function of $T_a$.}\label{fig:rhototcplotb}
	\end{subfigure}
~
	\begin{subfigure}[t]{0.475 \textwidth}
	\centering
	\includegraphics[width = \textwidth]{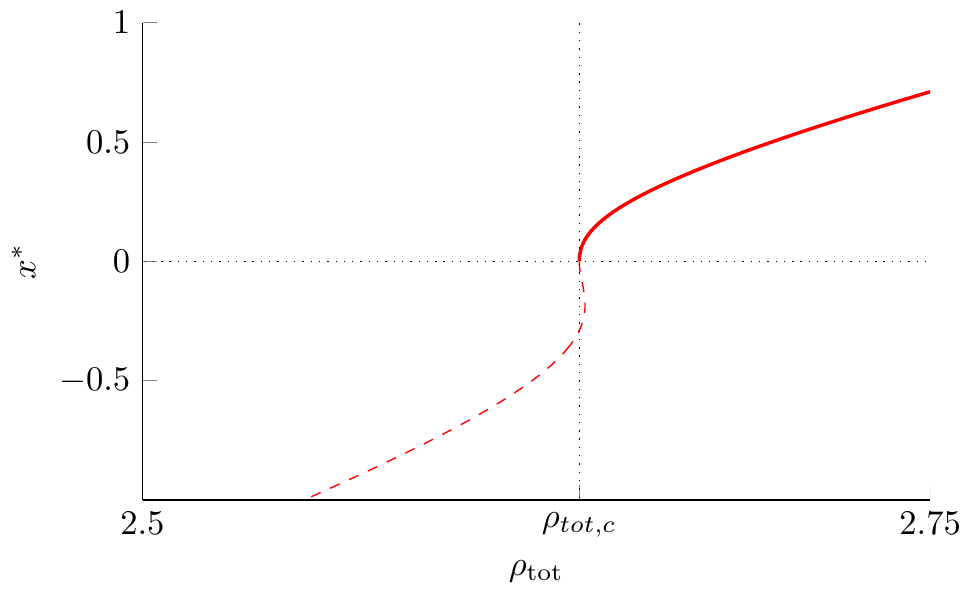}
	\caption{$x^*$ as function of $\rho_\mathrm{tot}$.}
	\end{subfigure}

	\caption{Plot of the critical colony size $\rho_{\mathrm{tot},c}$ as function of the ambient temperature $T_a$, according to equation~\eqref{eq:criticalbiomass} (a) and a plot of $x^*$ as function of $\rho_\mathrm{tot}$ for a type II solution. Parameter values used are $\bar{f} = 1$, $\chi_1 = 1$, $L = 10$ and $T_{\chi} = 25$; in (b) $T_a = 0$ and $\chi_2 = 1$.}
	\label{fig:rhototcplot}
\end{figure}

Note that in the above expressions,  the constant $C_1>0$ is still present. Hence, for different choices of $C_1$, we find different solutions. However, this does not lead to a solution for all $C_1>0$ because the condition $T_x=S(x)<0$ also needs to be satisfied for all $x \in [0,L]$. 
 Hence, the maximum temperature is achieved at $x = 0$, i.e. $T_{max} = T_\mathrm{core}$. Also, for a  type I solution  we assume that $T(x) < T_\chi$ for all $x \in [0,L]$, and hence, also $T_\mathrm{core} < T_{\chi}$. Therefore, this type I solution ceases to exist  when $T_\mathrm{core} = T_\chi$ and at that point the solution switches to become of  type II. Together with~\eqref{eq:typeITcore}, this leads to 
\begin{equation}
	T_\chi - T_a = \frac{2}{\chi_1} \log\left[ \cosh\left( \frac{\chi_1}{2} \sqrt{C_{1,c}} L \right)\right],
\end{equation}
where $C_{1,c}$ is the critical value for the parameter $C_1$ that leads to $T(0) = T_{\chi}$. Rewriting  this relation we obtain an expression for this critical value,
\begin{equation}
	\sqrt{C_{1,c}} = \frac{2}{L\chi_1} \arccosh\left( \exp\left[\frac{\chi_1}{2}(T_\chi - T_a)\right]\right) .
\end{equation}
In equation~\eqref{eq:typeIC1general} we have related the constant $C_1$ to the colony size via a boundary condition. We now substitute this critical value $C_{1,c}$ into~\eqref{eq:typeIC1general} to obtain the critical colony size
\begin{align}
	\rho_{\mathrm{tot},c} 
& = \frac{\sqrt{C_{1,c}} \tanh\left( \frac{\chi_1}{2} \sqrt{C_{1,c}} L \right) }{\bar{f}} \\
& = \frac{2}{\bar{f} L \chi_1} \arccosh\left( \exp\left[\frac{\chi_1}{2}(T_\chi - T_a)\right]\right) \tanh\left( \arccosh\left( \exp\left[\frac{\chi_1}{2}(T_\chi - T_a)\right]\right)  \right) \\
& = \frac{2}{\bar{f} L \chi_1} \sqrt{1 - \exp\left[ -\chi_1(T_\chi - T_a)\right]} \log\left( \sqrt{\exp\left[\chi_1(T_\chi-T_a)\right]-1}+\exp\left[\frac{\chi_1}{2}(T_\chi-T_a)\right]\right)
\label{eq:criticalbiomass}
\end{align}
A sketch of the critical colony size $\rho_{\mathrm{tot},c}$ as function of the ambient temperature $T_a$ is given in Figure~\ref{fig:rhototcplotb}.

Since bees produce heat, logically, the presence of more bees leads to a larger core temperature $T_\mathrm{core}$. Only when enough bees are present can $T_\mathrm{core} > T_\chi$ (and the steady state solutions becomes a type II configuration). The critical colony size $\rho_{\mathrm{tot},c}$ at which this happens depends on the ambient temperature $T_a$ as well as the heat production capabilities of the bees (modelled by $\bar{f}$). When $f$ is described by the step-function~\eqref{eq:definitionchi}, equation~\eqref{eq:criticalbiomass} does not specify the correct critical colony size. Using the same arguments that led to~\eqref{eq:criticalbiomass} one can derive that the correct description of $\rho_{\mathrm{tot},c}$ is given by
\begin{multline} 
 \rho_{\mathrm{tot},c} = \frac{2}{\chi_1 L} \ln\left( \sqrt{\exp\left[\chi_1(T_\chi-T_a)\right]-1}+\exp\left[\frac{\chi_1}{2}(T_\chi-T_a)\right]\right) \cdot \\ \left\{ \frac{1}{f_\mathrm{low}} \sqrt{1 - \exp\left(-\chi_1[T_\chi-T_a]\right)} + \left( \frac{1}{f_\mathrm{high}}-\frac{1}{f_\mathrm{low}}\right) \sqrt{1 - \exp\left(-\chi_1[T_\chi-T_f]\right)} \right\}.
\label{eq:rhototad}
\end{multline}

\subsection{Type II solutions}
Next, we analyse the type II steady state solutions which are characterised by the fact that    $\chi(T)$ does change sign. Hence, there is a point $x^* \in [0,L)$ such that $\chi(T) < 0$ for $x \in [0,x^*)$ and $\chi(T) > 0$ for $x \in (x^*,L]$. The construction of such a solution therefore entails solving both of these equations and matching them at $x = x^*$; that is, at this matching point we require $T$, $S$ and $S_x$ to be continuous.

\subsubsection{The first region where $x \in [0, x^*)$}

In the first region, $\chi(T)=-\chi_2 < 0$, and therefore, in a similar way as for the type I solution,  system~\eqref{eq:steadystateODE} reduces to
\begin{equation}
	0 = S_{xx} + \chi_2 S S_x.
\end{equation}
Following the same approach as before -- integrating this once -- yields
\begin{equation}
	S_x = - \frac{\chi_2}{2} \left(S^2 + D_1\right),
\end{equation}
where it follows from the boundary condition $S(0) = 0$ and the fact that $S_x(0) = - \bar{f} \rho(0) > 0$ that the constant $D_1$ must be positive. Then, the general solution to this  equation is
\begin{equation}
	S(x) = -\sqrt{D_1} \tan\left( \frac{\chi_2}{2} \sqrt{D_1} x + D_2 \right);  \hspace{1cm} \left(x < x^*\right).
	\label{eq:typeIIfirstpart}
\end{equation}
where the boundary condition $S(0) = 0$ enforces $D_2 = 0$. The constant $D_1$ will be determined when we match this solution to the solution in the second region.

\subsubsection{The second region where $x \in (x^*,L]$, and the matching of the two parts}

The solution in the second part has to satisfy the same equation~\eqref{eq:typeIODE} as the type I solution. The general solution to this equation is
\begin{equation}
	S(x) = - \sqrt{E_1} \tanh\left( \frac{\chi_1}{2} \sqrt{E_1} x + E_2 \right); \hspace{1cm} \left(x > x^*\right),
	\label{eq:typeIIsecondpart}
\end{equation}
where $E_1$ and $E_2$
are (so far) unknown  constants. A relation between these constants can be determined
by using  the boundary condition $S(L) = - \bar{f} \rho_\mathrm{tot}$. Moreover, matching
solution~\eqref{eq:typeIIfirstpart} to solution~\eqref{eq:typeIIsecondpart},
by assuming that $S$ and $S_x$ are continuous at 
  the switching point $x=x^*$,
  gives two additional relations between the constants $E_1$, $E_2$ and constant $D_1$ of the solution in the first part. This yields the following set of conditions:
\begin{align}
	\sqrt{E_1} \tanh\left( \frac{\chi_1}{2} \sqrt{E_1} L + E_2 \right) & = \bar{f} \rho_\mathrm{tot}; \label{eq:typeIIconstraint1}\\
	\sqrt{D_1} \tan\left( \frac{\chi_2}{2} \sqrt{D_1} x^*\right) & = \sqrt{E_1} \tanh\left( \frac{\chi_1}{2} \sqrt{E_1}x^* + E_2\right); \\
	\chi_2 D_1 \cos^{-2}\left( \frac{\chi_2}{2} \sqrt{D_1} x^*\right) & = 
\chi_1 E_1 \sech^2\left( \frac{\chi_1}{2} \sqrt{E_1}x^* + E_2\right).\label{eq:typeIIconstraint3}
\end{align}
A priori, it seems like the constants can be determined uniquely from these equations. However, the value $x^*$ for the switching point is still unknown and, more importantly, it depends on  the constants $D_1$, $E_1$ and $E_2$. Therefore, these three conditions do not form a closed system of equations. To find the  additional constraints, the temperature profile needs to be taken into account. This profile can be found by integrating $S$ once,
and is given by
\begin{equation}
	T(x) = 
	\begin{cases}
		T_\mathrm{core} + \frac{2}{\chi_2} \ln\left[ \cos\left( \frac{\chi_2}{2} \sqrt{D_1} x\right)\right], & \left(x < x^*\right);\\
		T_a + \frac{2}{\chi_1} \ln\left[\cosh\left( \frac{\chi_1}{2} \sqrt{E_1} L + E_2\right)\right] - \frac{2}{\chi_1} \ln\left[\cosh\left(\frac{\chi_1}{2} \sqrt{E_1}x + E_2\right)\right], & \left(x > x^*\right),
	\end{cases}
\end{equation}
where we have used the boundary condition $T(L) = T_a$ and the fact that $T(0)=T_\mathrm{core}$.
Now, $T$ needs to be continuous at $x=x^*$, and also $T(x^*) = T_\chi$, by definition.
This leads to the  two additional conditions,
\begin{align}
	T_\mathrm{core} + \frac{2}{\chi_2} \ln\left[ \cos\left( \frac{\chi_2}{2} \sqrt{D_1} x^*\right)\right] & = T_\chi; \label{eq:typeIIconstraint4}\\
	T_a + \frac{2}{\chi_1} \ln\left[\cosh\left( \frac{\chi_1}{2} \sqrt{E_1} L + E_2\right)\right] - \frac{2}{\chi_1} \ln\left[\cosh\left(\frac{\chi_1}{2} \sqrt{E_1}x^* + E_2\right)\right] & = T_\chi. \label{eq:typeIIconstraint5}
\end{align}
Finally, constraints~\eqref{eq:typeIIconstraint1}-\eqref{eq:typeIIconstraint3} and~\eqref{eq:typeIIconstraint4}-\eqref{eq:typeIIconstraint5} give five algebraic relations for the five unknown constants, $x^*$, $T_\mathrm{core}$, $D_1$, $E_1$ and $E_2$ which can be solved numerically. The  constants $x^*$ and $T_\mathrm{core}$ are given in Figures~\ref{fig:rhototcplot} and~\ref{fig:TcorePlots}, where $\rho_\mathrm{tot}$ and $T_a$ are varied. Note that in the region where the curves are dashed, $x^*$ becomes negative which is unrealistic and this solution does not exist there -- precisely when this happens a steady state changes from a type II to a type I configuration (or vice versa).

\begin{remark}
In this paper, we do not explicitly study the (spectral) stability of the stationary states constructed here -- see however Remark~\ref{energyFunctionalRemark} by which the stability of type I solutions can be settled. It is natural to expect that the (linear) stability of type II solutions can be (formally) settled by an approach similar to the present analysis.
\end{remark}

\section{The effect of a moving boundary}\label{sec:movingBoundary}
In this section, we study the extension of the model to a moving boundary instead of a fixed boundary at $x=L$. This is the setting in which the model originally was formulated in~\cite{watmough1995self}. Thus, we let the boundary of the colony be time-dependent, i.e. $L = L(t)$. The movement of this boundary needs to be such that, in the absence of mortality, no bees are created or lost, and hence, the colony size $\rho_\mathrm{tot}$ needs to remain constant. 
Therefore, we assume that $\frac{d}{dt}\rho_\mathrm{tot} =\frac{d}{dt} \int_0^{L(t)} \rho(x,t) dx = 0$ which results in the following equation for $L(t)$
\begin{equation}
 	\rho(L,t) L_t + \rho_x(L,t) + \chi(T(L,t))\rho(L,t) T_x(L,t) = 0.
	\label{eq:boundaryODE}
\end{equation}
Moreover, we need to replace the no-flux boundary condition at $x = L$. Following~\cite{watmough1995self}, we obtain a new boundary condition by defining the local density of bees at the end $x = L(t)$ of the colony to be constant $\rho_L$. Therefore, for the  system with moving boundary (\ref{eq:boundaryODE}) the full set of boundary conditions is
\begin{align}
	T_x(0,t) & = 0; & \rho_x(0,t) & = 0; \\
	T(L(t),t) & = T_a; & \rho(L(t),t) & = \rho_L. \label{eq:BCmovingBoundary}
\end{align}

\subsection{Stationary states in the system with a moving boundary}\label{sec:movingBC-steadyState}

The stationary solutions found in section~\ref{sec:steadyStates} are not influenced by the addition of a moving boundary to the model. Namely, to find steady state solutions of the model with a moving boundary, we need to set $\frac{dL}{dt} = 0$ in equation~\eqref{eq:boundaryODE}. This reduces this equation to the boundary condition,
\begin{equation*}
	\rho_x(L,t) + \chi(T(L)) \rho(L) T_x(L) = 0,
\end{equation*}
as before~\eqref{eq:boundc2}. Therefore, the steady state solutions of the model with a moving boundary are exactly the same as  those for the model with a fixed boundary. However, whereas previously we fixed the domain at a certain length $L$, this length $L$ now will be selected and still needs to be determined. For that we use the boundary condition~\eqref{eq:BCmovingBoundary} since the length $L$ is selected such that  $\rho(L) = \rho_L$. Using the expression for $\rho$ in~\eqref{eq:rhosrelation}, the fact that $T_x=S$ and expression~\eqref{eq:SsolI} yields  the following condition for a type I solution
\begin{equation}
	\frac{\chi_1}{2} C_1 \sech\left(\frac{\chi_1}{2} \sqrt{C_1} L\right)^2 = \bar{f} \rho_L,
\label{eq:relmovI}
\end{equation}
and for a type II solution, using~\eqref{eq:typeIIsecondpart},
to the condition
\begin{equation}
	\frac{\chi_1}{2} E_1 \sech\left(\frac{\chi_1}{2} \sqrt{E_1} L + E_2\right)^2 = \bar{f} \rho_L.
\label{eq:relmovII}
\end{equation}
Either of these conditions forms, together with the condition~\eqref{eq:typeIC1general} for type I, and~\eqref{eq:typeIIconstraint1}-\eqref{eq:typeIIconstraint3} and~\eqref{eq:typeIIconstraint4}-\eqref{eq:typeIIconstraint5} for type II solutions previously found,  a complete set of equations for all  the constants, that now also include the value of $L$.

\begin{figure}
		\centering
	\begin{subfigure}[t]{0.475\textwidth}
		\centering
		\includegraphics[width=\textwidth]{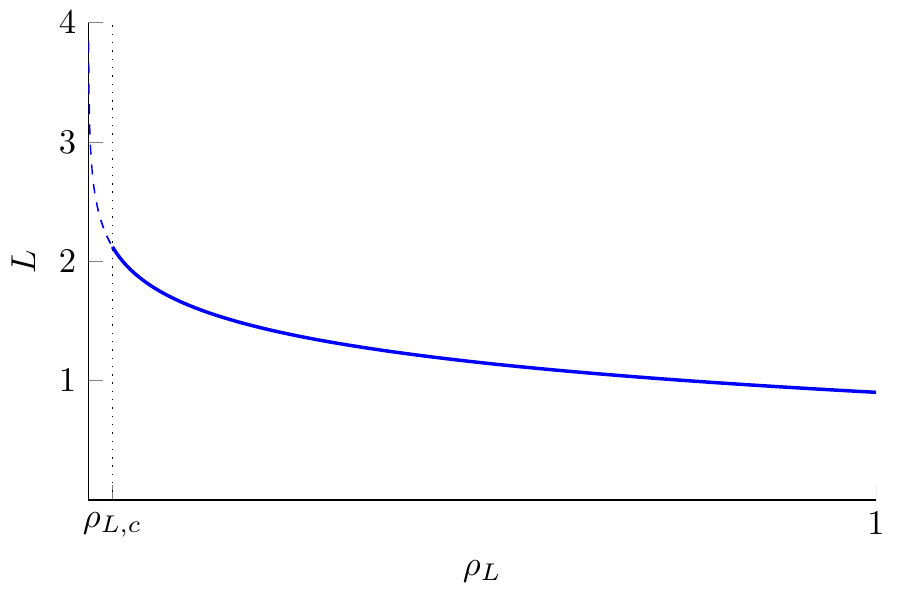}
		\caption{Steady state length $L$ as function of $\rho_L$}\label{fig:movBCtypeILa}
	\end{subfigure}
~
	\begin{subfigure}[t]{0.475\textwidth}
		\centering
		\includegraphics[width=\textwidth]{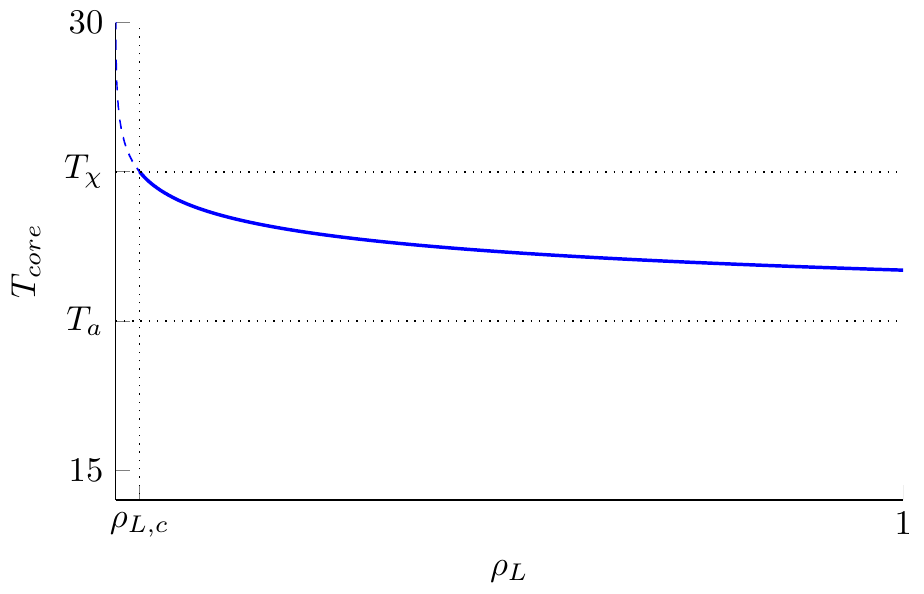}
		\caption{$T_\mathrm{core}$ as function of $\rho_L$}\label{fig:movBCtypeILb}
	\end{subfigure}

	\caption{Plots of the steady state length $L$ and the corresponding core temperature $T_\mathrm{core}$, for a type I solution on a domain with moving boundary, as function of $\rho_L$. The dotted lines in (b) denote $T_\mathrm{core} = T_\chi$ and $T_\mathrm{core} = T_a$, and $\rho_{L} = \rho_{L,c}$. The dashed blue part of the plots correspond to parameter combinations in which a type I solution does not exist, because $T_\mathrm{core} > T_{\chi}$. Parameters used are $\bar{f} = 1$, $\chi_1 = 1$, $\rho_\mathrm{tot} = 3$, $T_{a} = 20$, $T_{\chi} = 25$ and $\rho_\mathrm{tot} =3$.}
	\label{fig:movBCtypeIL}
\end{figure}

Interestingly, the constants for the type I solution can  be expressed in closed form
by taking the square of~\eqref{eq:typeIC1general} and adding this to~\eqref{eq:relmovI}:
\begin{align}
	C_1 & = \frac{2}{\chi_1} \bar{f} \rho_L + \bar{f}^2 \rho_\mathrm{tot}^2; \\
	L & = \frac{ 2 \arctanh\left( \bar{f} \rho_\mathrm{tot} / \sqrt{C_1} \right) }{\chi_1 \sqrt{C_1}} \label{eq:movBCtypeIL}
\end{align}
Using this, we plot $L$ as a function of  $\rho_L$ in Figure~\ref{fig:movBCtypeILa}. Clearly, $L$ decreases when $\rho_L$ increases. Moreover, $L \rightarrow \infty$ and $\sqrt{C_1}\rightarrow \bar{f} \rho_\mathrm{tot}$ when $\rho_L \downarrow 0$ and $L \rightarrow 0$ when $\rho_L \rightarrow \infty$. Similarly, the core temperature $T_\mathrm{core}$ in expression~\eqref{eq:typeITcore} follows the same pattern: $T_\mathrm{core} \rightarrow \infty$ when $\rho_L \downarrow 0$ and $T_\mathrm{core} \downarrow T_a$ when $\rho_L \rightarrow \infty$, see also Figure~\ref{fig:movBCtypeILb}.

\begin{figure}
	\begin{subfigure}[t]{0.475 \textwidth}
		\centering
		\includegraphics[width=\textwidth]{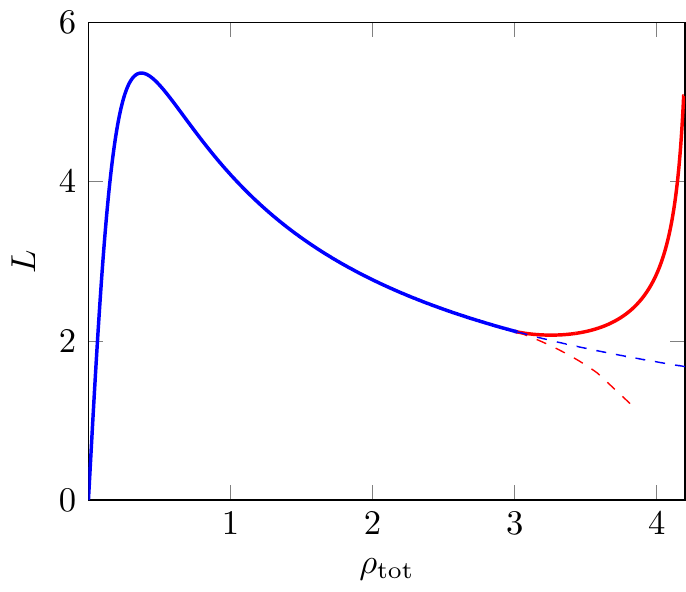}
		\caption{steady state $L$ as function of $\rho_\mathrm{tot}$}\label{fig:movBCtypeIIb}
	\end{subfigure}
~
	\begin{subfigure}[t]{0.475 \textwidth}
		\centering
		\includegraphics[width=\textwidth]{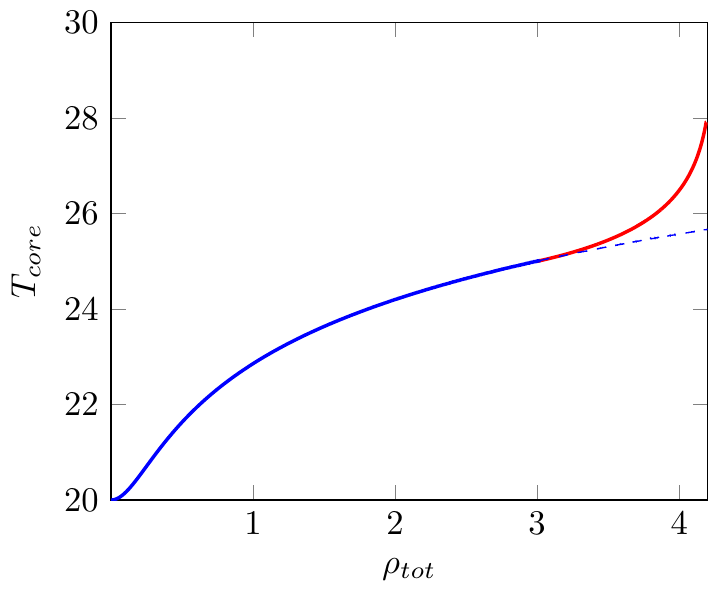}
		\caption{$T_\mathrm{core}$ as function of $\rho_\mathrm{tot}$}\label{fig:movBCtypeIIc}
	\end{subfigure}

	\caption{Plots of the steady state $L$ and $T_\mathrm{core}$ for a solution on a domain with moving boundary, as function of $\rho_\mathrm{tot}$. The blue lines indicate type I solutions, whereas the red lines indicate the type II solutions; dashed lines are solutions that are not realistic. Parameters used are $\bar{f} = 1$, $\chi_1 = 1$, $\chi_2 = 1$, $T_{\chi} = 25$ , $\rho_L = 0.0305$ and $T_a = 20$.}
	\label{fig:movBC}
\end{figure}

Since for type I solutions the temperature remains below $T_{\chi}$, this should also hold for the core temperature. Therefore, because $T_\mathrm{core}$ increases as $\rho_L \downarrow 0$, there exists a critical $\rho_{L,c}$: type I solutions only exist for $\rho_L > \rho_{L,c}$. In Figure~\ref{fig:movBC}, $L$ and $T_\mathrm{core}$ are also given as  functions of $\rho_\mathrm{tot}$. Here we see a generalisation of the results on a fixed domain; there is a critical colony size $\rho_{\mathrm{tot},c}$ below which type I solutions can exist and above which only type II solutions can exist.

For a type II solution the constraints~\eqref{eq:relmovII},~\eqref{eq:typeIIconstraint1}-\eqref{eq:typeIIconstraint3} and~\eqref{eq:typeIIconstraint4}-\eqref{eq:typeIIconstraint5} do not lead to any closed form expression. With the aid of a root-finding algorithm, we can find $L$ as a function of  $\rho_\mathrm{tot}$, as well as the relation between $T_\mathrm{core}$ and $\rho_\mathrm{tot}$ , see Figure~\ref{fig:movBC}. Comparing these to the plots on a domain with fixed boundary, i.e. Figure~\ref{fig:TcorePlots}, one sees similar behaviour. 

The most prominent difference between the model with and the model without a moving boundary is the computational difficulty in handling the moving boundary. To obtain realistic values for the steady state lengths $L$, the value of  $\rho_L$ needs to be chosen very small. This makes it difficult to determine a solution of the above constraints and leads to subtleties in the numerical simulations. Moreover, the boundary condition $\rho(L) = \rho_L$ is quite artificial. Especially when the mortality is added this is unrealistic, as this boundary condition forces a fixed local bee density at $x = L$, while the colony size decreases. When $\rho_\mathrm{tot} \downarrow 0$ this leads to sudden rapid changes in $L$ and unrealistic bee density profiles. Hence, in the set-up of~\cite{watmough1995self}, the moving boundary model is not adequate to study bee losses. For the study of the model with mortality we therefore decided to stick to the most simple model with constant length $L$ to avoid these problems.

\section{The effect of mortality}\label{sec:mortality}

In sections~\ref{sec:steadyStates} and~\ref{sec:movingBC-steadyState} we  studied the steady state solutions of the bee model~\eqref{eq:beePDE} without mortality, i.e. $\theta = 0$. We showed in those sections that for each value for the colony size, $\rho_\mathrm{tot}$, there exists a steady state. We will denote this steady state by $(T_s,\rho_s)(x;\rho_\mathrm{tot})$. Simulations indicate that this is the only steady state and that it is the only attractor of the system; for a type I solution this can be proven using an energy functional, see Remark~\ref{energyFunctionalRemark}.

In this section we add mortality of bees to the system, i.e. $\theta \neq 0$. Because the bees die on a much longer time scale than that their  movement happens, we assume the mortality coefficient  to be small. Therefore, the amount of bees, $\rho_\mathrm{tot}$, only decreases  slowly and the bees have enough time to rearrange themselves. This means that the system can be approximated by the so-called quasi-stationary states given by $(T_s,\rho_s)(x;\rho_\mathrm{tot}(t))$, where the only time-dependence comes from the (slow) change in $\rho_\mathrm{tot}$.

To find the evolution of $\rho_\mathrm{tot}(t) = \int_0^L \rho(x,t)\ dx$, we determine its derivative using~\eqref{eq:beePDE} with boundary conditions, as
\begin{equation}
	\frac{d \rho_\mathrm{tot}}{dt}(t) = -\int_0^L \theta\left(\rho(x,t),T(x,t)\right)\ \rho(x,t)\ dx.
\label{eq:biomassEvolution}
\end{equation}
Note that the system conserves mass if $\theta = 0$, i.e. when there is no mortality.

The mortality rate $\theta$ as formulated in the introduction, in~\eqref{eq:mort}, thus leads to
\begin{equation}
	\frac{d \rho_\mathrm{tot}}{dt}(t) = - \frac{\theta_0}{\rho_\mathrm{tot}(t)^\gamma} \left(1 + \frac{m}{\rho_\mathrm{tot}(t)}\right) \int_0^L \mathbbm{1}_{T(x,t) < T_\theta}\ \rho(x,t)^2\ dx, \label{eq:biomassEvolutionSpecific}
\end{equation}
where $\mathbbm{1}$ is the indicator function. Even in the present most simplified setting, this expression is (too) hard to fully study analytically. In section~\ref{sec:simulations} we therefore use numerical simulations to study the evolution of $\rho_{tot}$. However, it is possible to use asymptotic analysis to determine what happens when $\rho_\mathrm{tot}$ is small (under the quasi-steady state assumption).

\begin{figure}[t!]
	\centering
	\begin{subfigure}[t]{0.18\textwidth}
		\centering
		\includegraphics[width=\textwidth]{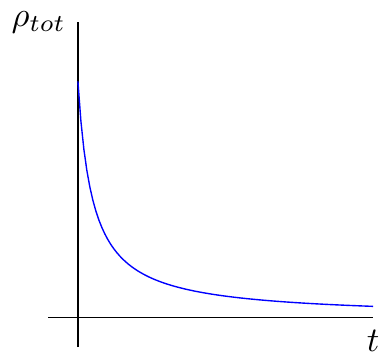}
		\caption{\footnotesize $m = 0$, $\gamma < 1$;\\$m \neq 0$, $\gamma < 0$}
	\end{subfigure}
~
	\begin{subfigure}[t]{0.18\textwidth}
		\centering
		\includegraphics[width=\textwidth]{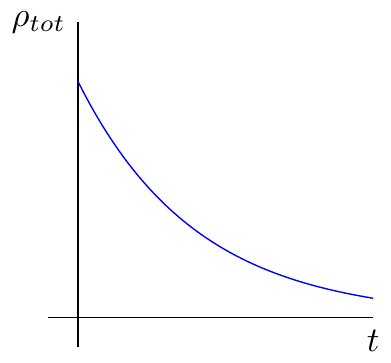}
		\caption{\footnotesize $m = 0$, $\gamma = 1$;\\$m \neq 0$, $\gamma = 0$}
	\end{subfigure}
~
	\begin{subfigure}[t]{0.18\textwidth}
		\centering
		\includegraphics[width=\textwidth]{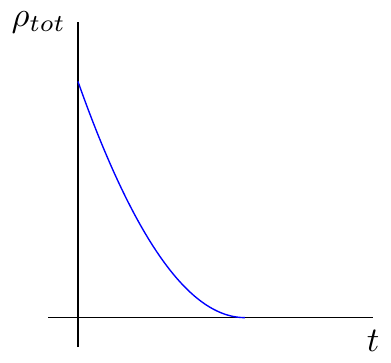}
		\caption{\footnotesize $m = 0$,$\gamma \in (1,2)$;\\$m \neq 0$, $\gamma \in (0,1)$}
	\end{subfigure}
~
	\begin{subfigure}[t]{0.18\textwidth}
		\centering
		\includegraphics[width=\textwidth]{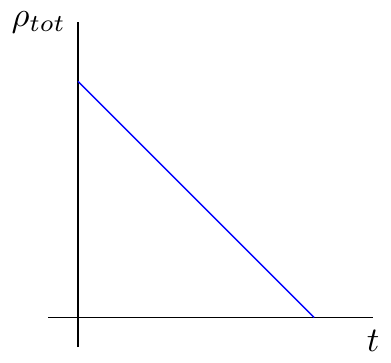}
		\caption{\footnotesize $m = 0$, $\gamma = 2$;\\$m \neq 0$, $\gamma = 1$}
	\end{subfigure}
~
	\begin{subfigure}[t]{0.18\textwidth}
		\centering
		\includegraphics[width=\textwidth]{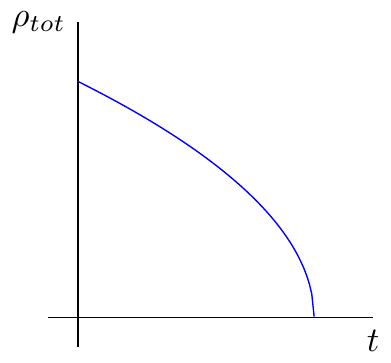}
		\caption{\footnotesize $m = 0$, $\gamma > 2$;\\$m \neq 0$, $\gamma > 1$}
		\label{fig:rhototODEsolutionalphaM1}
	\end{subfigure}

\caption{The qualitative different possible solutions to the colony size evolution ODE~\eqref{eq:biomassEvolutionSpecificSmall} in the absence of mites ($m = 0$) and in the presence of mites ($m > 0$).}
\label{fig:rhototODEsolutions}
\end{figure}

Specifically, we set $\rho_\mathrm{tot} = \varepsilon \tilde{\rho}_\mathrm{tot}$, where $0 < \varepsilon \ll 1$ and $\tilde{\rho}_\mathrm{tot} = \mathcal{O}(1)$ with respect to $\varepsilon$. Since $\varepsilon$ is small, $\rho_\mathrm{tot} < \rho_{\mathrm{tot},c}$ in this case and the steady state configuration is of type I. Hence from~\eqref{eq:typeIC1general} we obtain $C_1 = \mathcal{O}(\varepsilon)$. So we set $C_1 = \varepsilon \tilde{C}_1$ and using Taylor approximations obtain
\begin{equation}
	\tilde{C}_1 =  \frac{2 \bar{f} \tilde{\rho}_\mathrm{tot}}{\chi L} + \mathcal{O}(\varepsilon^2). \label{eq:C1Tilde}
\end{equation}
Then, by~\eqref{eq:rhoDistributionTypeI} we have
\begin{equation}
	\rho(x,t) = \varepsilon \frac{\tilde{\rho}_{\mathrm{tot}}(t)}{L} + \mathcal{O}(\varepsilon^2).
\end{equation}
Also, \eqref{eq:typeITcore} along with the expansion~\eqref{eq:C1Tilde} reveals $T(x) < T_\mathrm{core} < T_\theta$ since $T_\theta > T_a$. Hence, for small $\rho_\mathrm{tot}$ the evolution of $\rho_\mathrm{tot}$ in~\eqref{eq:biomassEvolutionSpecific} is to leading order given by
\begin{align}
	\frac{d \rho_\mathrm{tot}}{dt}(t) &= - \frac{\theta_0}{\varepsilon^\gamma \tilde{\rho}_\mathrm{tot}(t)^\gamma} \left(1 + \frac{m}{\varepsilon \tilde{\rho}_\mathrm{tot}(t)}\right) \int_0^L \left( \frac{\varepsilon^2 \tilde{\rho}_{\mathrm{tot}}(t)^2}{L^2} + \mathcal{O}(\varepsilon^3) \right) \ dx \nonumber \\ 
& = - \frac{\theta_0}{L} \frac{ m + \varepsilon \tilde{\rho}_\mathrm{tot}(t)}{\varepsilon^{\gamma-1} \tilde{\rho}_\mathrm{tot}(t)^{\gamma-1}} + \mathcal{O}\left( m \varepsilon^{2-\gamma} + \varepsilon^{3-\gamma} \right). \label{eq:biomassEvolutionSpecificSmall}
\end{align}
Thus, if $m \neq 0$ we have
\begin{align}
	\frac{d \tilde{\rho}_\mathrm{tot}}{dt} & = - \frac{m \theta_0}{L} \varepsilon^{-\gamma} \tilde{\rho}_\mathrm{tot}(t)^{1-\gamma} + \mathcal{O}(\varepsilon^{1-\gamma}),
\end{align}
and if $m = 0$ we have
\begin{align}
	\frac{d \tilde{\rho}_\mathrm{tot}}{dt} & = - \frac{\theta_0}{L} \varepsilon^{1-\gamma} \tilde{\rho}_\mathrm{tot}(t)^{2-\gamma} + \mathcal{O}(\varepsilon^{2-\gamma}),
\end{align}
As a consequence, $\tilde{\rho}_\mathrm{tot}$ is to leading order given by
\begin{equation}
	\tilde{\rho}_\mathrm{tot}(t) = 
	\begin{cases}
		\tilde{\rho}_\mathrm{tot}(0)e^{-\frac{\theta_0}{L}t}, & \mbox{ if $m = 0$ and $\gamma = 1$}; \\
		\left[ \tilde{\rho}_\mathrm{tot}(0)^{\gamma-1} - (\gamma-1) \frac{\theta_0}{L}\varepsilon^{1-\gamma} t \right]^{\frac{1}{\gamma-1}}, & \mbox{ if $m = 0$ and $\gamma \neq 1$};\\
		\tilde{\rho}_\mathrm{tot}(0) e^{-\frac{\theta_0 m}{L} t}, & \mbox{ if $m \neq 0$ and $\gamma = 0$};\\
		\left[ \tilde{\rho}_\mathrm{tot}(0)^\gamma - \gamma \frac{\theta_0 m}{L} t \varepsilon^{-\gamma}\right]^{\frac{1}{\gamma}}, & \mbox{ if $m \neq 0$ and $\gamma \neq 0$}.
	\end{cases}
\end{equation}
In Figure~\ref{fig:rhototODEsolutions} the qualitative different possible evolutions of $\rho_\mathrm{tot}(t)$ are shown based on the parameters $\gamma$ and $m$. Clearly, the presence of mites has an amplifying effect on the decline of bees in a colony. When mites are present ($m \neq 0$) the death rate is of higher order compared to a colony without mites ($m = 0$). Thus (the last) bees in colonies with mites are expected to die faster than those in colonies without.

\section{Simulations}\label{sec:simulations}
In this section, we give the results of the numerical simulations we performed on system~\eqref{eq:beePDE} with the mortality rate given in~\eqref{eq:mort}. As one of the key goals of this research is to better understand colony deaths during the winter period, we are interested in the evolution of the colony size, $\rho_\mathrm{tot}(t)$. Because the model only includes mortality of bees and no birth of new bees -- since bees do not reproduce in winter -- $\rho_\mathrm{tot}(t)$ will decrease over time, and $\lim_{t \rightarrow \infty} \rho_\mathrm{tot}(t) = 0$. Moreover, simulations indicate (almost) all bees have died after a finite (extinction) time $t_e$. Numerically we can obtain the value $t_e$ such that $\rho_\mathrm{tot}(t_e) \approx 0$. To overcome the winter period, a colony needs to survive until the new flowering season in spring; that is, the time to extinction needs to satisfy $t_e > t_f$, where $t_f$ is the time between start and end of the winter season ($t_f \approx 2 \cdot 10^6$ minutes; $5$ months). Thus, it is of great interest to determine how the extinction time $t_e$ is prolonged or shortened by the various causes of death. By the complexity of the differential equation for $\rho_\mathrm{tot}(t)$, equation~\eqref{eq:biomassEvolutionSpecific}, we resort to numerical simulations.

For the simulations in this section, we have chosen parameter values that are relatively realistic. No major changes were observed in simulations with other parameter combinations (provided that $T_\theta < T_\chi$). In this section, we have used functions $f$ and $\chi$ that were adapted from ~\cite{watmough1995self}, and were simplified to be piecewise constants. More specifically, we have used
\begin{equation}
	f(T) = 
	\begin{cases}
		3, & \mbox{if $T < 15$};\\
		0.6, & \mbox{if $T \geq 15$};
	\end{cases}
\hspace{1cm}
	\chi(T) =
	\begin{cases}
		1, & \mbox{if $T < 25$};\\
		-1, & \mbox{if $T \geq 25$}.
	\end{cases}
\end{equation}
Little information was available on the parameters present in the mortality term $\theta$. The only data we are aware of, stipulates that a healthy colony (i.e. one with no mites; $m = 0$) with ambient temperature $T_a = 0$ and initial size\footnote{In reality, colonies typically have around $10000$ to $15000$ bees. Since we model a cross-section of a colony, the value for $\rho_{tot}(0)$ does not match with those values; instead, we have used~\cite{watmough1995self} to determine the typical value of $\rho_{tot}(0)$ for a cross-section; in this way, the used values for $\rho_{tot}(0)$ correspond to colonies with a realistic amount of bees.} of $\rho_\mathrm{tot}(0) = 10$ loses roughly half of the colony in $100$ days ($\approx 1.5 \cdot 10^5$ minutes)~\cite{dooremalen2012}. Because of this we have tuned the values for $\gamma$, $T_\theta$ and $\theta_0$ to be in line with these measurements. Ultimately, this led to two choices: (i) $\gamma = 1$, $T_\theta = 21$, $\theta_0 = 4 \cdot 10^{-3}$ and (ii) $\gamma = 2$, $T_\theta = 21$, $\theta_0 = 4 \cdot 10^{-2}$. For other exponents $\gamma \geq 1$ results are expected to be similar.

\begin{figure}
		\centering
	\begin{subfigure}[t]{0.475\textwidth}
	\centering
		\includegraphics[width = \textwidth]{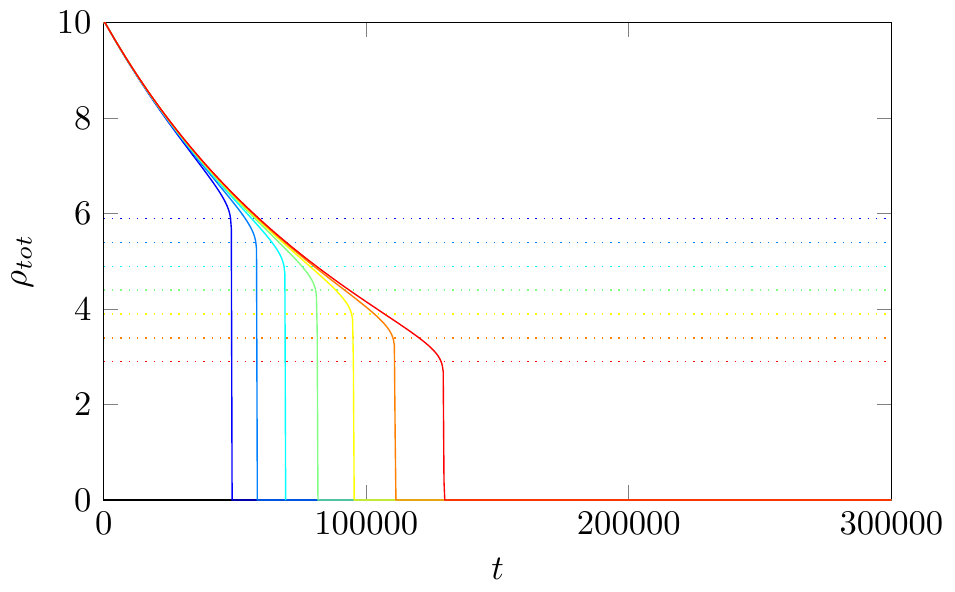}
	\caption{$\gamma = 1$}
	\end{subfigure}
~
	\begin{subfigure}[t]{0.475\textwidth}
	\centering
		\includegraphics[width =\textwidth]{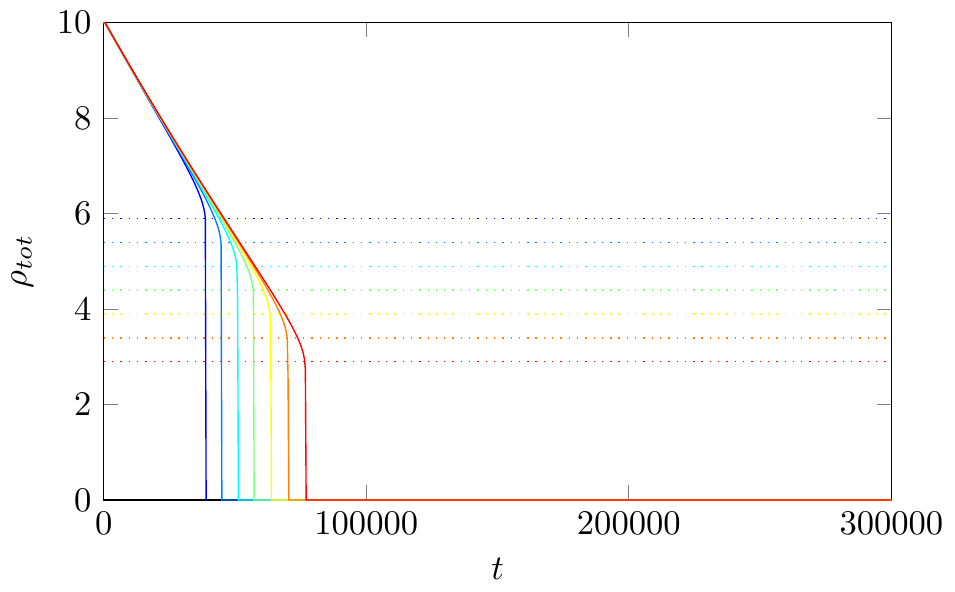}
	\caption{$\gamma = 2$}
	\end{subfigure}
		\caption{The evolution of $\rho_\mathrm{tot}(t)$ for $m = 20$, $\rho_\mathrm{tot}(0) = 10$ and $T_a$ varied from $T_a = -9$ (dark blue) to $T_a = +9$ (red) with increments of $3$ for simulations with $\gamma = 1$ (a) and simulations with $\gamma = 2$ (b). The dotted lines indicate the critical colony size $\rho_{\mathrm{tot},c}$ below which the solution is of type I (recall that $\rho_{\mathrm{tot},c}$ depends on $T_a$). The simulations in this figure show that $T_a$ crucially influences the critical colony size $\rho_{\mathrm{tot},c}$ and therefore the moment at which the speed-up in the decrease of $\rho_\mathrm{tot}(t)$ sets in; as long as $\rho_\mathrm{tot}(t) > \rho_{\mathrm{tot},c}$ the precise value of $T_a$ does not alter the evolution much.}
		\label{fig:simulationDataTa}
\end{figure}

\begin{figure}
	\centering
	\begin{subfigure}[t]{0.475\textwidth}
	\centering
		\includegraphics[width=\textwidth]{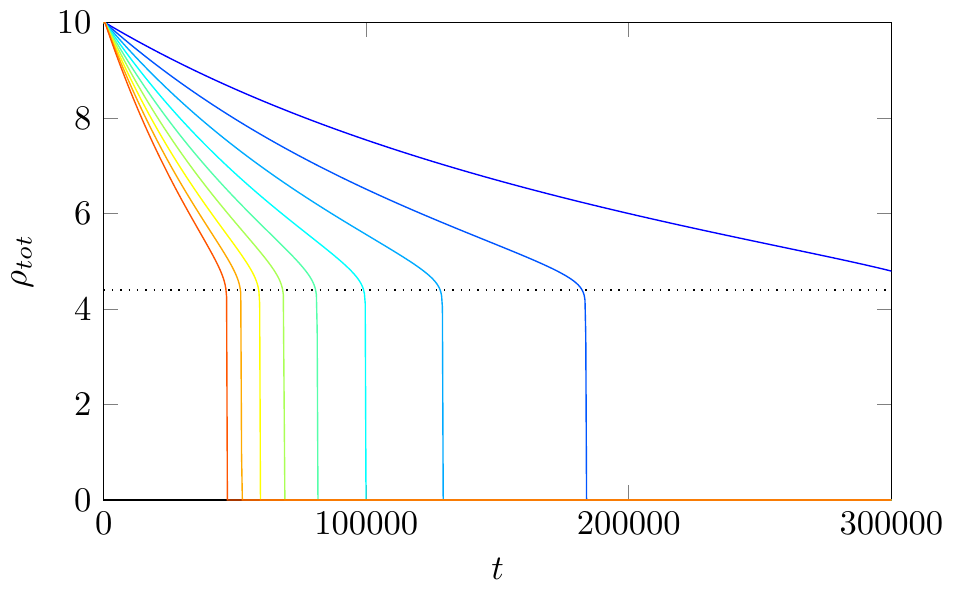}
	\caption{$\gamma = 1$}
	\end{subfigure}
~
	\begin{subfigure}[t]{0.475\textwidth}
	\centering
		\includegraphics[width=\textwidth]{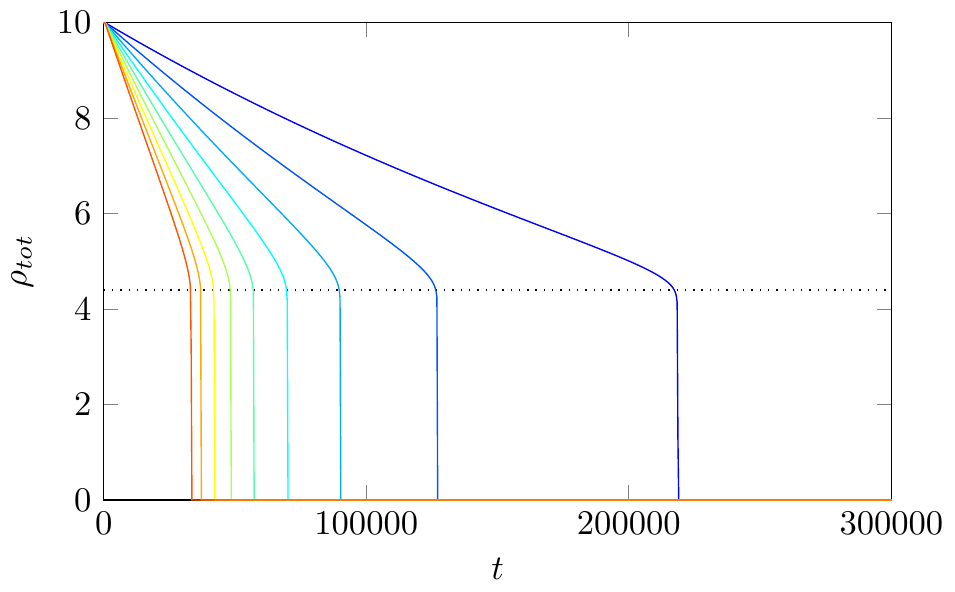}
	\caption{$\gamma = 2$}
	\end{subfigure}		
		\caption{The evolution of $\rho_\mathrm{tot}(t)$ for $T_a = 0$, $\rho_\mathrm{tot}(0) = 10$ and $m$ varied between $m = 0$ (dark blue) and $m = 40$ (red) with increments of $5$ for simulations with $\gamma = 1$ (a) and simulations with $\gamma = 2$ (b). The dotted black line indicates the critical colony size $\rho_{\mathrm{tot},c}$ below which the solution is of type I. The simulations show the impact of $m$ on the survival of individual bees -- which is particularly large for small $m$ values, but is diminished for larger $m$ values.}
		\label{fig:simulationDataM}
\end{figure}

\begin{figure}
	\centering
	\begin{subfigure}[t]{0.475\textwidth}
	\centering
		\includegraphics[width=\textwidth]{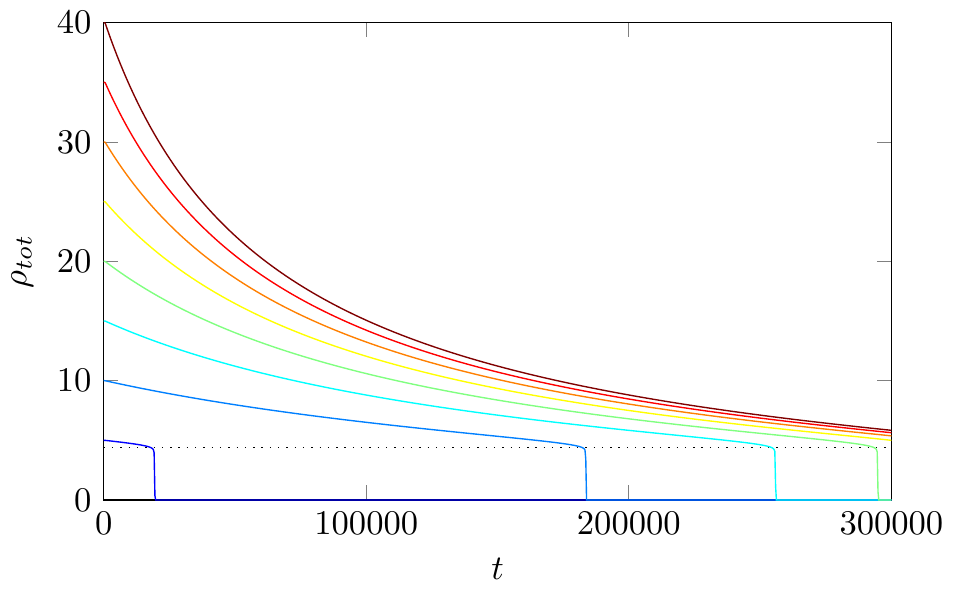}
	\caption{$\gamma = 1$}
	\end{subfigure}
~
	\begin{subfigure}[t]{0.475\textwidth}
	\centering
		\includegraphics[width=\textwidth]{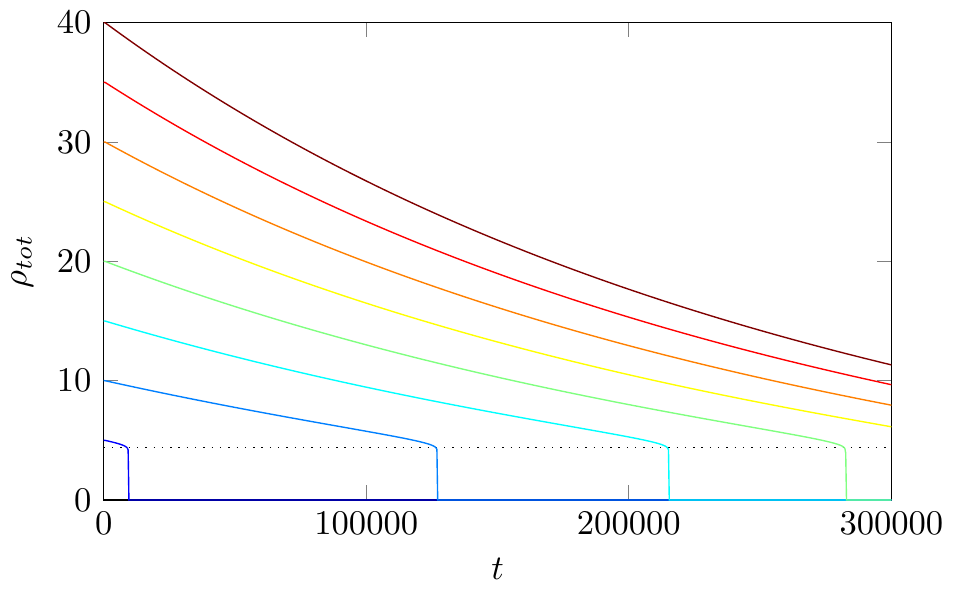}
	\caption{$\gamma = 2$}
	\end{subfigure}			
		\caption{The evolution of $\rho_\mathrm{tot}(t)$ for $m = 5$, $T_a = 0$ and $\rho_\mathrm{tot}(0)$ ranging from $\rho_\mathrm{tot}(0) = 5$ (dark blue) and $\rho_\mathrm{tot}(0) = 40$ (red) with increments of $5$ for simulations with $\gamma = 1$ (a) and simulations with $\gamma = 2$ (b). The black dotted line indicates the critical colony size $\rho_{\mathrm{tot},c}$. These simulations show the effect of increasing the initial colony size, which leads to improvements for the survival of the colony. These improvements are large for smaller initial colony sizes, but become smaller for larger colony sizes.}
		\label{fig:simulationDataB}
\end{figure}

In this section, we vary the parameters $m$ (amount of mites in the colony), $T_a$ (the ambient temperature) and $\rho_\mathrm{tot}(0)$ (the initial colony size), to determine their impact on the colony size evolution $\rho_\mathrm{tot}(t)$ in general, and the extinction time $t_e$ specifically. From our numerical simulations (see Figures~\ref{fig:simulationDataTa},~\ref{fig:simulationDataM},~\ref{fig:simulationDataB}), a sudden speed-up in colony size decline can be seen after which the colony quickly dies out. Upon better inspection, we have determined that this speed-up happens when the colony configuration changes from type II to type I (see Figure~\ref{fig:intro-steadyStateTypes} and Figure~\ref{fig:steadystatesolutions}). That is, when the colony size  decreases below the critical size $\rho_{\mathrm{tot},c}$, the colony is unable to survive much longer. In the simulation results (see Figures~\ref{fig:simulationDataTa},~\ref{fig:simulationDataM},~\ref{fig:simulationDataB}) we  have indicated the critical colony size $\rho_{\mathrm{tot},c}$. In section \ref{sec:steadyStates}, we derived an expression for $\rho_{\mathrm{tot},c}$, see~\eqref{eq:rhototad}. Note that $\rho_{\mathrm{tot},c}$ is influenced (only) by the ambient temperature $T_a$ (and not by the amount of mites $m$, or initial colony size $\rho_\mathrm{tot}(0)$).

In order to perform the numerical simulations, we have discretized~\eqref{eq:beePDE} to an ODE, which we have solved using MATLAB's ode15s function. Next, we give the results of the numerical simulations when varying $T_a$, $m$ and $\rho_\mathrm{tot}(0)$.

\paragraph{The effect of the ambient temperature $T_a$}

First, we vary the ambient temperature $T_a$ between $T_a=-9$ (dark blue) to $T_a=9$ (red)
while taking $m=20$ and $\rho_\mathrm{tot}(0)=10$. We give the results in Figure~\ref{fig:simulationDataTa}. We also denote the critical colony size $\rho_{\mathrm{tot},c}$ 
by dotted black lines. This is the critical value below which the solution is of type I. 
When $T_a$ is smaller, $\rho_{\mathrm{tot},c}$ is larger, see also Figure~\ref{fig:rhototcplot}
of which an analog can be given for~\eqref{eq:rhototad}.
We observe that, 
since $T_a$ influences the critical colony size, $\rho_{\mathrm{tot},c}$, a change in $T_a$ directly effects the survival of individual bees in a colony. Therefore, the extinction time $t_e$ is smaller when $T_a$ is decreased. Ecologically, this means that more bees are needed to create enough heat for the colony in harsher winters.

The ambient temperature, however, does not have a significant impact on the evolution of $\rho_\mathrm{tot}(t)$ when bees still form a type II configuration. Therefore, especially the ambient temperature towards the end of the winter period is crucial for the survival of a colony. If it is very cold at the start of the winter, this is predicted to have only little impact, whereas a cold end of the winter can dramatically decrease the extinction time $t_e$.

\paragraph{The effect of the amount of mites in a bee colony $m$}

Next, we vary the influence of the mites by increasing the parameter $m$ from 
$m=0$ (no mites; dark blue) to $m=40$ (red). The simulations for $T_a=0$ and  $\rho_\mathrm{tot}(0) = 10$ are given in Figure~\ref{fig:simulationDataM}.
When a colony is pestered by lots of mites, individual bees during winter may have a lower body condition~\cite{amdam2004, vandooremalen2013,blanken2015}. Therefore, when there are more  mites  in a colony (larger $m$), bees die faster as is indeed observed in the figure. However, the amount of mites does not play any role in the value of $\rho_{\mathrm{tot},c}$. Therefore mites mainly influence the rate of 
the  death of bees  for type II configurations; when there are  more mites,   a healthy type II solution degrades faster into an unhealthy type I configuration. Moreover, the simulations indicate that mites have a significant impact on the survival of the colony. Especially the transition from a colony with no mites to a colony with a few mites has a huge impact; for a very large number of mites the effect of adding the same number of mites on the extinction time $t_e$ is diminished.

\paragraph{The effect of initial colony size $\rho_\mathrm{tot}(0)$}

We also ran several simulations for different initial colony sizes, $\rho_\mathrm{tot}(0)$; see Figure~\ref{fig:simulationDataB}. We choose $T_a=0$, $m=5$ and vary 
$\rho_\mathrm{tot}(0)$ between $\rho_\mathrm{tot}(0)=5$ (dark blue) to $\rho_\mathrm{tot}(0)=10$ (red)
 Clearly, having more bees in a colony ensures that the colony survives for a longer period, because $\rho_\mathrm{tot}(t) > \rho_{\mathrm{tot},c}$ for a longer time period. However, simply increasing the initial colony size does not necessarily increase the extinction time, $t_e$ by much: when $\rho_\mathrm{tot}(0)$ is close to the critical size $\rho_{\mathrm{tot},c}$, a larger initial colony size indeed leads to a vast improvement. However, for larger  $\rho_\mathrm{tot}(0)$, the effect of increasing the initial colony size does not lead  to  a much longer survival time. For instance, in Figure~\ref{fig:simulationDataB}, the difference in the time to extinction between simulations with $\rho_\mathrm{tot}(0) = 5$ and $\rho_\mathrm{tot}(0) = 10$ is enormous compared to the difference between those with $\rho_\mathrm{tot}(0) = 15$ and $\rho_\mathrm{tot}(0) = 20$.

\section{Multiple honey combs}\label{sec:simulations-multipleRows}

In the previous sections, we have inspected the model~\eqref{eq:beePDE} that describes the evolution of $T$ and $\rho$ in the space between two honey combs. However, in reality, bees use multiple honey comb and the colony divides itself in parts, with each part clumping together in one inter-comb space when it gets too cold and they have to stick together to create enough heat to warm the hive. Bee connect these parts by moving through and around the combs separating the inter-comb spaces. It is possible to extend the model~\eqref{eq:beePDE} to include multiple of these inter-comb spaces. As a short encore, in this section we briefly discuss how, and present a simple example of such an extended model.

A straightforward way to extend the model~\eqref{eq:beePDE} is to copy the model for each inter-comb space and then add movement of bees between the colony parts. Specifically, an extended model (for $N$ inter-comb spaces) can have the following form
\begin{equation}
\begin{cases}
	\frac{d T_i}{dt} & = \frac{d^2 T_i}{dx^2} + f(T_i) \rho_i + \sum_{j \neq i} I_{ij}^T(T_i,T_j,\rho_i,\rho_j), \\
	\frac{d \rho_i}{dt} & = \frac{d^2 \rho_i}{dx^2} - \frac{d}{dx} \left[ \chi(T_i) \rho_i \frac{d T_i}{dx}\right] - \theta(\rho_i,T_i) + \sum_{j \neq i} I_{ij}^\rho(T_i,T_j,\rho_i,\rho_j).
\end{cases}
\hspace{0.5cm} (i \in \{1, \ldots, N\})
\end{equation}
Here, $T_i$ and $\rho_i$ denote temperature respectively local bee density in the $i$-th inter-comb space. The functions $I_{ij}^T$ and $I_{ij}^\rho$ should be constructed such that they capture the effect of interactions between combs, e.g. bee movement between combs.

Multiple choices are possible to model these interaction terms. Here, we refrain from going into the details of the nature of interactions between bees and temperature in the case of multiple combs. Instead, and as an example, we -- once again -- choose a strongly simplified set-up: if the core temperature of a colony part gets below a critical temperature $T_c$ -- this temperature needs to be chosen above, but close to the temperature $T_\chi$ -- bees in that inter-comb space try to escape to warmer inter-combs spaces at a certain rate depending on the difference between core temperatures. Specifically, we have chosen the interaction functions
\begin{align*}
	I_{ij}^T & = 0; && (i,j \in \{1, \ldots, N\}) \\
	I_{ij}^\rho & = - \mathbbm{1}_{T{i,core} < T_c}\ \alpha (T_{j,core} - T_{i,core}) \rho_i\ +\ \mathbbm{1}_{T_{j,core} < T_c}\ \alpha (T_{i,core} - T_{j,core}) \rho_j, && (i,j \in \{1, \ldots, N\})
\end{align*}
where $\mathbbm{1}$ is the indicator function, and $\alpha$ a parameter that measures the movement rate of bees (per temperature difference between inter-comb spaces).

\begin{figure}
	\centering
		\includegraphics[width=0.475\textwidth]{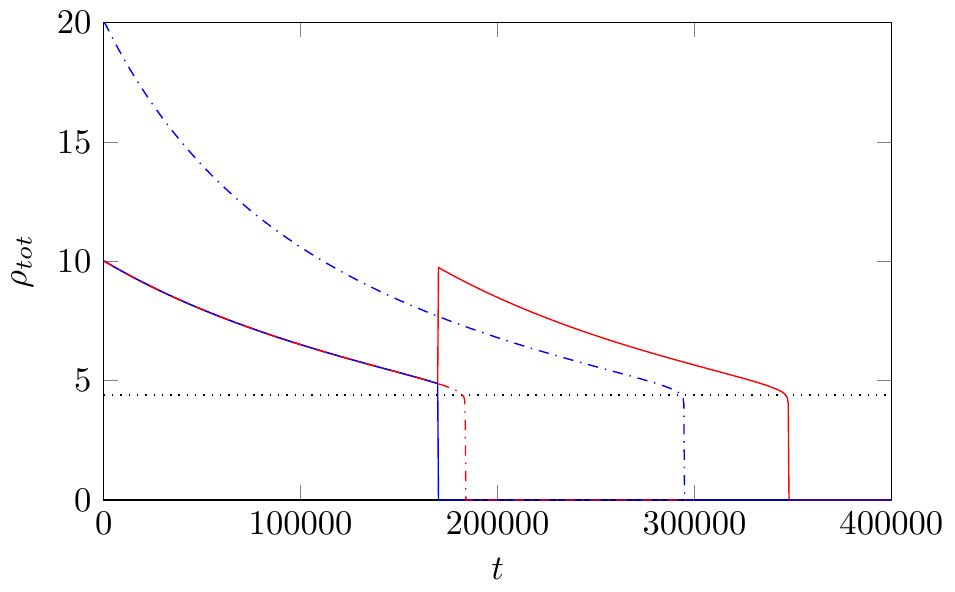}
	\caption{Evolution of $\rho_\mathrm{tot}(t)$ in a simulation with two colony parts, each occupying the space between two combs (one in red, one in blue). Simultaneously shown are two simulations with the whole colony in one inter-comb space (dashed-dotted lines), one of which started with $\rho_\mathrm{tot}(0) = 20$ (blue) and one started with $\rho_\mathrm{tot}(0) = 10$ (red). The dotted black lines indicate the critical biomass $\rho_{\mathrm{tot},c}$ below which the solution is of type I. For these simulations we have used the parameters $T_a = 0$, $m = 5$, $\gamma = 1$ and $\alpha = 1$.}
	\label{fig:multipleCombs}
\end{figure}

In Figure~\ref{fig:multipleCombs}, we show a simulation of a colony that consists of two parts occupying spaces between combs, both of which are initially occupied with the same number of bees, i.e. $\rho_{1,tot}(0) = \rho_{2,tot}(0) = 10$. At first, both colony parts lose roughly the same amount of bees due to mortality, similar to the case with the whole colony in one inter-comb space. Then, when the core temperate of a colony part decreases below $T_c$, the remaining bees evacuate one of the inter-comb spaces -- the colder one -- and move to the other, that is (relatively) warmer. As a result, one inter-comb space is left with no bees and the other one has all of them. The rest of the simulation then continues as if it was a simulation with the whole colony in one inter-comb space.

From Figure~\ref{fig:multipleCombs}, it is clear why this is beneficial for a bee colony. By evacuating one inter-comb space, the bees can cluster together and make sure only one colony part dies -- and not both. The figure also indicates why bee colonies use multiple combs in reality in the first place -- it helps to improve the survival of the colony: in simulations, colonies occupying two inter-comb spaces have a longer extinction-time $t_e$ compared to colonies with the same initial total number of bees clustered in only one inter-comb space; for instance, in Figure~\ref{fig:multipleCombs} clearly having two colony parts with $\rho_\mathrm{tot}(0)=20$ leads to a larger $t_e$ than having one with $\rho_\mathrm{tot}(0) = 40$ does.

This idea extends to colonies with more combs. However, here it becomes important to stipulate how bees move between the inter-comb spaces that are warm enough (i.e. those with $T_{i,\mathrm{core}} > T_c$). As, in principle, bees can move freely between inter-comb spaces, it seems logical that `quite soon' they have distributed themselves equally over the warmer inter-comb spaces -- justifying the initial equal bee distribution in the simulations with two inter-comb spaces in Figure~\ref{fig:multipleCombs}. This idea could be modelled by a diffusive motion between inter-comb spaces -- though for this work we have not studied this in (any) detail. All in all, the simulation of Figure~\ref{fig:multipleCombs} clearly shows that allowing for (bee) dynamics between inter-comb spaces may prolong the extinction time and thus improve a colony's survival. Choosing the optimal movement strategy and (mathematically) understanding the associated mechanism is the subject of ongoing research.

\section{Discussion}\label{sec:conclusions}

In this article, we postulated and analysed a (highly) simplified model for honey bee colonies in winter. During this winter period, mortality of individual honey bees is high, especially in the presence of parasites, and linked to a colony's capability to generate and preserve heat. Therefore, thermoregulation is the most important process in the colony to understand, which is modelled as a combined effect of heat generating shivering and an individual thermotactic movement towards a preferred temperature $T_\chi$. This leads to a model of Keller-Segel type, in which the chemotactic coefficient $\chi$ changes sign at the bees preferred temperature $T_\chi$.

Because of this change of sign in the chemotactic coefficient $\chi$, the model (in absence of mortality) possesses two different types of steady state solutions, depending on the amount of bees $\rho_{tot}$ in a colony. When a colony has too little individuals, a colony's core temperature does not exceed the preferred temperature and bees cluster together with the highest density in the core; alternatively, when enough bees are in a colony, the core temperature does exceed the preferred temperature and bees form an (isolating) band at the edge of the colony -- in agreement with observations~\cite{stabentheiner2003,stabentheiner2010}. Other recent models were developed to simulate thermoregulation in honey bee clusters (see for example~\cite{eskov2009mathematical,ocko2014collective}). These models do not explore the consequences of individual bee mortality on colony survival and the decrease in thermoregulatory ability of a colony. Hence they do not explore the role of colony size and how changes in colony size affect thermoregulation during winter.

We added mortality to the model of honey bee colonies during winter. Since mortality of bees in winter has been linked to the amount of work a bee has to perform~\cite{amdam2004}, we postulated that mortality is influenced by (i) the effect of local temperature, (ii) the effect of effective refresh rates allowing recovery after a bout of generating heat and (iii) the effect of parasitic mites in a colony. As mortality is important on a long timescale compared to bee movement, a colony closely follows the mentioned steady states, with $\rho_{tot}$ acting as a slowly decreasing parameter. In simulations, we observed a sudden rapid decline in the amount of bees when the colony's form changed from type II to type I, i.e. when $\rho_{tot}$ decreased below the critical value. This suggests the rapid decline is related to the failing of a bee colony's theremoregulation when too little bees are left in a colony. For now, it remains unclear precisely how this rapid decline sets in from the mathematical point of view; inferring its origin forms an interesting avenue for further mathematical research.

For a colony to survive winter, (enough) bees need to survive until the new flowering season in spring. Therefore, it is important to understand how long a colony can survive in winter and how that is related to colony size at the start of winter. Our simulations support the following findings. The colder the ambient temperature (especially towards the end of winter), the more bees are needed to keep the colony in the type II form, and thus the sooner a colony is expected to die when the size becomes smaller than the critical threshold. Moreover, when mites are present, more bees die because of the elevated mortality rate, and a colony thus collapses earlier. Finally, when a colony has more bees, it is expected to survive longer. However, the correspondence between a colony's size and its expected lifetime is highly nonlinear and follows the law of diminishing returns. Future studies could investigate the relation between colony size before winter and survival changes at the beginning of the next spring.

In this context, it is also of interest to understand the behaviour of bees in a colony with multiple combs. In this article, we briefly considered a simple model where the bee colony is divided over two inter-comb spaces; simulations show that a (good) distribution of bees over multiple inter-comb spaces enhances the colony's survival and improves its expected lifetime. A careful analysis and modelling of more refined multi-comb models might be illuminating; it is expected that such analysis reveals the most optimal way to distribute bees over multiple inter-comb spaces, and might indicate why bees produce multiple combs in reality.

The model studied in this article has deliberately been chosen to be as simple as possible to allow for explicit mathematical analysis. However, results are expected to hold for more complicated models, including those with more realistic functions for e.g. the chemotactic movement and a bee's heat generation. Specifically, the increase in mortality due to failed thermoregulation should persist in more realistic models. Therefore, it is interesting to explore how a colony's thermoregulation can be optimized with the aim to improve a colony's winter survival time. During the year, bee keepers should reduce the mite load in the colony~\cite{vandooremalen2013}. Moreover, they have the opportunity just before winter starts to increase the size of the colony by eventually merging small colonies that are expected to die during winter. 

From a more mathematical perspective, there are also a lot of new research opportunities. The introduction of a chemotactic coefficient that changes sign leads to new, different behaviour compared to the (Keller-Segel type) models typically considered in the mathematical literature, such as the presence of two different types of steady-state solutions. Classical methods do no longer suffice in this context and novel methods need to be developed -- even to infer (global) stability of these steady state solutions in absence of mortality. There are also additional lines of research possible in case mortality is present. A significant next step would be a mathematical analysis of the impact of a slowly varying mortality. This would embed the present study in the strongly evolving research field of pattern formation under slowly varying circumstances, that also has a very direct relevance within developmental biology and studies on the effect of climate change (see~\cite{crampin1999reaction, crampin2002pattern, neville2006interactions, tzou2015slowly, bastiaansen2019dynamics} and references therein).

\section*{Acknowledgements}
We thank Jan Just Keijser and the National Institute for Subatomic Physics Nikhef for providing computing power to run the numerical simulations. R.B was supported by a grant within the Mathematics of Planet Earth program of the Netherlands Organization of Scientific Research (NWO).

\bibliographystyle{plain}
\bibliography{sources}

\begin{thebibliography}{10}

\bibitem{amdam2004}
Gro~V Amdam, Klaus Hartfelder, Kari Norberg, Arne Hagen, and Stig~W Omholt.
\newblock Altered physiology in worker honey bees (hymenoptera: Apidae)
  infested with the mite varroa destructor (acari: Varroidae): a factor in
  colony loss during overwintering?
\newblock {\em Journal of economic entomology}, 97(3):741--747, 2004.

\bibitem{amdam2009}
Gro~V Amdam, Olav Rueppell, M~Kim Fondrk, Robert~E Page, and C~Mindy Nelson.
\newblock The nurse’s load: Early-life exposure to brood-rearing affects
  behavior and lifespan in honey bees (apis mellifera).
\newblock {\em Experimental gerontology}, 44(6-7):467--471, 2009.

\bibitem{amdam2002}
Gro~Vang Amdam and Stig~W Omholt.
\newblock The regulatory anatomy of honeybee lifespan.
\newblock {\em Journal of theoretical biology}, 216(2):209--228, 2002.

\bibitem{bastiaansen2019dynamics}
Robbin Bastiaansen and Arjen Doelman.
\newblock The dynamics of disappearing pulses in a singularly perturbed
  reaction--diffusion system with parameters that vary in time and space.
\newblock {\em Physica D: Nonlinear Phenomena}, 388:45--72, 2019.

\bibitem{bellomo2015}
Nicola Bellomo, Abdelghani Bellouquid, Youshan Tao, and Michael Winkler.
\newblock Toward a mathematical theory of keller--segel models of pattern
  formation in biological tissues.
\newblock {\em Mathematical Models and Methods in Applied Sciences},
  25(09):1663--1763, 2015.

\bibitem{blanchet2015}
Adrien Blanchet, Jos{\'e}~Antonio Carrillo, David Kinderlehrer, Micha{\l}
  Kowalczyk, Philippe Lauren{\c{c}}ot, and Stefano Lisini.
\newblock A hybrid variational principle for the keller--segel system in r2.
\newblock {\em ESAIM: Mathematical Modelling and Numerical Analysis},
  49(6):1553--1576, 2015.

\bibitem{blanken2015}
Lisa~J Blanken, Frank van Langevelde, and Coby van Dooremalen.
\newblock Interaction between varroa destructor and imidacloprid reduces flight
  capacity of honeybees.
\newblock {\em Proc. R. Soc. B}, 282(1820):20151738, 2015.

\bibitem{crampin1999reaction}
Edmund~J Crampin, Eamonn~A Gaffney, and Philip~K Maini.
\newblock Reaction and diffusion on growing domains: scenarios for robust
  pattern formation.
\newblock {\em Bulletin of mathematical biology}, 61(6):1093--1120, 1999.

\bibitem{crampin2002pattern}
EJ~Crampin, WW~Hackborn, and PK~Maini.
\newblock Pattern formation in reaction-diffusion models with nonuniform domain
  growth.
\newblock {\em Bulletin of mathematical biology}, 64(4):747--769, 2002.

\bibitem{doke2015}
Mehmet~Ali Doeke, Maryann Frazier, and Christina~M Grozinger.
\newblock Overwintering honey bees: biology and management.
\newblock {\em Current Opinion in Insect Science}, 10:185--193, 2015.

\bibitem{esch1960}
Harald Esch.
\newblock {\"U}ber die k{\"o}rpertemperaturen und den w{\"a}rmehaushalt von
  apis mellifica.
\newblock {\em Zeitschrift f{\"u}r vergleichende Physiologie}, 43(3):305--335,
  1960.

\bibitem{esch1964}
Harald Esch.
\newblock Beitrfige zum {P}roblem der {E}ntfernungsweisung in den
  {S}chwfinzeltfinzen der {H}onigbienen.
\newblock {\em Zeitschrift f{\"u}r vergleichende Physiologie}, 48:534--546,
  1964.

\bibitem{eskov2009mathematical}
EK~Eskov and VA~Toboev.
\newblock Mathematical modeling of the temperature field distribution in insect
  winter clusters.
\newblock {\em Biophysics}, 54(1):85--89, 2009.

\bibitem{gallai2009}
Nicola Gallai, Jean-Michel Salles, Josef Settele, and Bernard~E Vaissi{\`e}re.
\newblock Economic valuation of the vulnerability of world agriculture
  confronted with pollinator decline.
\newblock {\em Ecological economics}, 68(3):810--821, 2009.

\bibitem{hayes2008}
Jerry Hayes~Jr, Robyn~M Underwood, Jeffery Pettis, et~al.
\newblock A survey of honey bee colony losses in the us, fall 2007 to spring
  2008.
\newblock {\em PloS one}, 3(12):e4071, 2008.

\bibitem{heinrich1981}
Bernd Heinrich.
\newblock Energetics of honeybee swarm thermoregulation.
\newblock {\em Science}, 212(4494):565--566, 1981.

\bibitem{heinrich2013}
Bernd Heinrich.
\newblock {\em The hot-blooded insects: strategies and mechanisms of
  thermoregulation}.
\newblock Springer Science \& Business Media, 2013.

\bibitem{hillen2009}
Thomas Hillen and Kevin~J Painter.
\newblock A user’s guide to pde models for chemotaxis.
\newblock {\em Journal of mathematical biology}, 58(1-2):183, 2009.

\bibitem{horstmann2003}
D~Horstmann.
\newblock From 1970 until present: The keller-segel model in chemotaxis and its
  consequences i.
\newblock 105:103--165, 01 2003.

\bibitem{keller1971}
Evelyn~F Keller and Lee~A Segel.
\newblock Model for chemotaxis.
\newblock {\em Journal of theoretical biology}, 30(2):225--234, 1971.

\bibitem{klein2007}
Alexandra-Maria Klein, Bernard~E Vaissiere, James~H Cane, Ingolf
  Steffan-Dewenter, Saul~A Cunningham, Claire Kremen, and Teja Tscharntke.
\newblock Importance of pollinators in changing landscapes for world crops.
\newblock {\em Proceedings of the Royal Society of London B: Biological
  Sciences}, 274(1608):303--313, 2007.

\bibitem{moritz2012}
Robin Moritz and Edward~E Southwick.
\newblock {\em Bees as superorganisms: an evolutionary reality}.
\newblock Springer Science \& Business Media, 2012.

\bibitem{neumann2010}
Peter Neumann and Norman~L Carreck.
\newblock Honey bee colony losses.
\newblock {\em Journal of Apicultural Research}, 49(1):1--6, 2010.

\bibitem{neville2006interactions}
AA~Neville, PC~Matthews, and HM~Byrne.
\newblock Interactions between pattern formation and domain growth.
\newblock {\em Bulletin of mathematical biology}, 68(8):1975--2003, 2006.

\bibitem{ocko2014collective}
Samuel~A Ocko and L~Mahadevan.
\newblock Collective thermoregulation in bee clusters.
\newblock {\em Journal of The Royal Society Interface}, 11(91):20131033, 2014.

\bibitem{southwick1983}
Edward~E Southwick.
\newblock The honey bee cluster as a homeothermic superorganism.
\newblock {\em Comparative biochemistry and physiology Part A: Physiology},
  75(4):641--645, 1983.

\bibitem{southwick1971}
Edward~E Southwick and John~N Mugaas.
\newblock A hypothetical homeotherm: the honeybee hive.
\newblock {\em Comparative Biochemistry and Physiology Part A: Physiology},
  40(4):935--944, 1971.

\bibitem{stabentheiner2010}
Anton Stabentheiner, Helmut Kovac, and Robert Brodschneider.
\newblock Honeybee colony thermoregulation--regulatory mechanisms and
  contribution of individuals in dependence on age, location and thermal
  stress.
\newblock {\em PLoS One}, 5(1):e8967, 2010.

\bibitem{stabentheiner2003}
Anton Stabentheiner, Helga Pressl, Thomas Papst, Norbert Hrassnigg, and Karl
  Crailsheim.
\newblock Endothermic heat production in honeybee winter clusters.
\newblock {\em Journal of Experimental Biology}, 206(2):353--358, 2003.

\bibitem{tautz2008}
J{\"u}rgen Tautz.
\newblock {\em The buzz about bees: biology of a superorganism}.
\newblock Springer Science \& Business Media, 2008.

\bibitem{tindall2008}
Marcus~J Tindall, Philip~K Maini, Steven~L Porter, and Judith~P Armitage.
\newblock Overview of mathematical approaches used to model bacterial
  chemotaxis ii: bacterial populations.
\newblock {\em Bulletin of mathematical biology}, 70(6):1570, 2008.

\bibitem{tzou2015slowly}
Justin~C Tzou, Michael~J Ward, and Theodore Kolokolnikov.
\newblock Slowly varying control parameters, delayed bifurcations, and the
  stability of spikes in reaction--diffusion systems.
\newblock {\em Physica D: Nonlinear Phenomena}, 290:24--43, 2015.

\bibitem{vandooremalen2013}
C~Van~Dooremalen, E~Stam, L~Gerritsen, B~Cornelissen, J~Van~der Steen,
  F~Van~Langevelde, and T~Blacqui{\`e}re.
\newblock Interactive effect of reduced pollen availability and varroa
  destructor infestation limits growth and protein content of young honey bees.
\newblock {\em Journal of insect physiology}, 59(4):487--493, 2013.

\bibitem{dooremalen2012}
Coby van Dooremalen, Lonne Gerritsen, Bram Cornelissen, Jozef~JM van~der Steen,
  Frank van Langevelde, and Tjeerd Blacqui{\`e}re.
\newblock Winter survival of individual honey bees and honey bee colonies
  depends on level of varroa destructor infestation.
\newblock {\em PloS one}, 7(4):e36285, 2012.

\bibitem{watmough1995self}
James Watmough and Scott Camazine.
\newblock Self-organized thermoregulation of honeybee clusters.
\newblock {\em Journal of Theoretical Biology}, 176(3):391--402, 1995.

\end{thebibliography}


\end{document}